\documentclass[12pt,preprint]{aastex6}

\usepackage{color}
\usepackage{mathrsfs}

\def\VEC#1{\mbox{\boldmath $#1$}}

\slugcomment{}

\shorttitle{Generalized GRMHD equations for plasmas 
in the EHT era}
\shortauthors{Koide}

\begin{document}


\title{Generalized general-relativistic magnetohydrodynamic equations 
for plasmas of active galactic nuclei in the era of the Event Horizon Telescope}


\author{Shinji Koide}
\affil{Department of Physics, Kumamoto University, 
2-39-1, Kurokami, Kumamoto, 860-8555, JAPAN}
\email{koidesin@kumamoto-u.ac.jp}



\begin{abstract}
The generalized general-relativistic magnetohydrodynamic (generalized GRMHD) equations
have been used to study specific relativistic-plasmas phenomena,
such as relativistic magnetic reconnection or wave propagation
modified by non-ideal MHD effects.
However, the $\Theta$ term in the generalized 
Ohm law, which expresses the energy exchange
between two fluids composing a plasma,
has yet to be determined in these equations.
In this paper, we determine the $\Theta$ term based on the
generalized relativistic Ohm law itself.
This provides closure of the generalized GRMHD equations, yielding a closed
\textcolor{black}{system} of the equations of relativistic plasma.
According to this \textcolor{black}{system} of equations, 
we reveal the characteristic scales of non-ideal MHD phenomena and
clarify the applicable condition of the ideal GRMHD equations.
We evaluate the characteristic scales of
the non-ideal MHD phenomena in the M87* plasma using the Event Horizon
Telescope observational data.
\end{abstract}



\section{Introduction} \label{sec1}

The observation of the supermassive black-hole shadow at the center
of the giant elliptical galaxy M87 with impressible images
by the Event Horizon Telescope (EHT) Collaboration \citep{eht19a}
has brought us into a new era of black-hole plasma physics.
To extract detailed information concerning the dynamics of the plasma, 
the black hole's gravitational field, and the black hole itself,
it is necessary to develop fully general-relativistic models of the accretion flow, 
associated winds and relativistic jets, as well as the emission properties 
of the plasmas.
The most popular approach to modeling dynamic relativistic sources 
is known as
the ``ideal general relativistic magnetohydrodynamic (GRMHD) approximation".
Over the last past few decades, a number of ideal GRMHD codes have been developed
and applied to a large variety of astrophysical scenarios
\citep{koide98,koide99,koide00,koide02,koide03,koide04,koide06,
gammie03,mckinney06,delzanna07,mckinney09,mckinney13,radice13}.
The EHT team also found that the images produced by GRMHD simulations 
with general-relativistic ray-tracing calculations \citep{eht19b,porth19}
are consistent with the asymmetric ring feature seen in the EHT data.
Comparing the GRMHD simulations and
the EHT images, the EHT Collaboration team concluded 
that the brightness asymmetry in the ring can be explained 
by the relativistic beaming of the emission from plasma rotating
close to the speed of light around a black hole spinning clockwise \citep{eht19b}.

In these ideal GRMHD simulations, 
both the finite electrically resistive effect, 
as well as a number of plasma effects, such as
the Hall effect, thermo-electromotive force, and electron inertia, are neglected.
These neglected effects may modify the dynamics of the plasma and the magnetic field
in processes like magnetic reconnection. Actually,
\citet{acciari09} presented simultaneous radio and $\gamma$-ray observations
of M87* and showed that radio knots were ejected from the core of the galaxy
where the TeV $\gamma$-ray flare occurred.
This knot may be recognized by the plasmoids formed by the magnetic
reconnection like coronal-mass ejection from the Sun.
\citet{hirota13,hirota15} and \citet{comisso14} 
showed that electron inertia
causes collisionless magnetic reconnection.

To investigate the specific properties of relativistic plasmas with non-ideal MHD effects, 
we must use generalized 
GRMHD, including a generalized relativistic version of Ohm law.
The generalized GRMHD equations were introduced on the basis of the two-fluid approximation
of plasma in the Kerr metric in a pioneering study by \citet{khanna98}.
More generalized equations from the general-relativistic Vlasov--Boltzmann
equation in time-varying space-time were formulated by \citet{meier04}.
\citet{koide08} introduced some peculiar quantity-average definitions 
to derive the generalized special-relativistic MHD equations for pair plasma
from the relativistic two--fluid equations without any additional approximations. 
\citet{koide09} extended these generalized special-relativistic MHD equations 
for pair plasma to
any two-component plasma, including 
electron--ion (normal) plasma. 
The general-relativistic version of generalized MHD for any kind of
plasma, including pair and normal plasmas, was given by \citet{koide10}.
Recently, non-ideal MHD effects
(e.g., related to relativistic magnetic reconnection
and wave propagation)
have been analyzed by a number of authors
using the generalized relativistic MHD equations
\citep{asenjo15a,asenjo15b,kawazura17,kawazura_etal17,yang16,yang18,yang17,yang19a,yang19b,yang19c,liu18a,liu19}.
The equations of the generalized relativistic MHD are identical
to the relativistic equations of the two-fluid approximation
with specific averages of physical variables.
However, the $\Theta$ term in the generalized Ohm law 
was not determined in the previous papers \citep{koide08,koide09,koide10}.
This $\Theta$ term describes energy transport from the negatively charged fluid
to the positively charged one.
To determine the $\Theta$ term,
\citet{koide09} assumed that the relative velocity of the two fluids is not so large that
the frictional force is proportional to the relative velocity
 and obtained
\begin{equation}
\Theta =  \frac{\theta}{2e\rho_{\rm e}'} 
\frac{(\rho_{\rm e}'^2+J_\nu J^\nu) \left \{\Delta \mu 
\left [ n^2 - \Delta \mu n \frac{\rho_{\rm e}'}{e} - \mu 
\left ( \frac{\rho_{\rm e}'}{e} \right )^2 \right ]  
+n \frac{\rho_{\rm e}'}{e} \right \} }
{\left (n+ \frac{\Delta \mu}{2} \frac{\rho_{\rm e}'}{e} \right ) 
\left [ n^2 - \Delta \mu n \frac{\rho_{\rm e}'}{e} - \mu 
\left ( \frac{\rho_{\rm e}'}{e} \right )^2 \right ]  
}  ,
\label{eot}
\end{equation}
where 
the definitions of variables $e$, $\mu$, $\Delta \mu$, $n$, $J^\mu$, and $\rho'_{\rm e}$
are shown in Section \ref{sec2}.
Unfortunately, the $\Theta$ term cannot be determined via Equation (\ref{eot}) 
because it contains the unknown parameter, $\theta$ ($0 \le \theta \le 1$).

In this paper, we determine the $\Theta$ term in the generalized Ohm law
using the covariant form of the generalized relativistic Ohm law itself.
Previously, to determine the $\Theta$ term, we had considered a number of approaches, 
including the relativistic Vlasov--Boltzmann
equation and the collisional two-fluid approximation; however,
we finally found that such efforts are not necessary to determine the $\Theta$ term.
Using this term, we obtain an explicitly closed \textcolor{black}{system} of the generalized GRMHD equations.

Using the closed \textcolor{black}{system} of generalized GRMHD equations, 
we determine the characteristic scales at which non-ideal MHD effects,
such as the resistive electromotive-force effect,
the Hall effect, the thermo-electromotive force, and the current-carrier (electrons,
in the normal plasma case) inertia effect become significant 
in the generalized Ohm law. For this purpose, we introduce some plasma
parameters from linear analysis of special-relativistic plasma waves.

Using the parameters of plasmas obtained from observations by the EHT Collaboration,
we can evaluate the characteristic scales of the non-ideal MHD phenomena of
the plasmas around M87*.
The evaluated scales show that in global phenomena at the scale of the horizon radius,  
the additional terms of the Hall effect, thermo-electromotive force,
and electron inertia including the electric resistivity
are negligible. Thus, the ideal GRMHD is regarded as a good approximation
of the global dynamics of the plasmas around M87*.

In Section \ref{sec2}, we review the generalized GRMHD equations 
based on the general-relativistic two-fluid equations.
We derive the $\Theta$ term of the generalized Ohm law by itself.
In Section \ref{sec4}, we \textcolor{black}{briefly show}
the 3 + 1 formalism of the generalized GRMHD equations 
with the normal observer frame.
In Section \ref{sec5}, we introduce some plasma parameters obtained by linear analysis of
relativistic plasma waves and show the characteristic scales of the non-ideal MHD
phenomena of the relativistic plasmas.
We evaluate the significance of the non-ideal MHD terms
of the generalized GRMHD equations in Section \ref{sec6} 
using the plasma parameters introduced
in Section \ref{sec5} and reveal the characteristic scales of the non-ideal MHD effects.
We apply these characteristic scales to the plasma around the black hole
of M87*
using observational data from the EHT Collaboration \citep{eht19b} in the last part of Section 
\ref{sec5}. 
The final section presents a summary of this paper.

\section{Generalized GRMHD equations}
 \label{sec2}
\subsection{Review of the generalized GRMHD equations}

We review the generalized GRMHD equations based on the general relativistic 
two-fluid equations \citep{koide10}.
For simplicity, we assumed that the plasma is composed of two fluids,
where one fluid consists of positively charged particles with
mass $m_+$ and electric charge $e$ and the other fluid consists
of negatively charged particles with mass $m_-$ and electric charge $-e$.
We take no account of radiation cooling effect, plasma viscosity, and 
self-gravity in order to study the fundamentals of interaction
between magnetic fields and resistive plasmas around the spinning
black holes. We also assumed that the plasmas are heated 
only by Ohmic heating and disregarded nuclear reactions, pair creation,
and annihilation. 
The space-time, $(x^0,x^1,x^2,x^3)=(t,x^1,x^2,x^3)$, is characterized
by a metric $g_{\mu \nu}$, where a line element is given by
$ds^2 = g_{\mu\nu} dx^\mu dx^\nu$. Here, we use units in which
the speed of light, the dielectric constant, and the magnetic 
permeability in vacuum all are unity: $c=1$, $\epsilon_0=1$, $\mu_0=1$.
When we consider a black hole with the mass $M_{\rm BH}$, we use the unit system
such as $G M_{\rm BH} = 1$, where $G$ is the gravitational constant.
The relativistic equations of the two fluids and the Maxwell equations
are 
\begin{eqnarray} 
\nabla_\nu (n_\pm U_\pm^\nu) &=& 0 , \label{4formnum} \\ 
\nabla_\nu (h_\pm U_\pm^\mu U_\pm^\nu)  &=& 
-\nabla^\mu p_\pm \pm e n_\pm g^{\mu\sigma} U_\pm^\nu F_{\sigma\nu} 
\pm R^\mu , \label{4formmom} \\ 
\nabla_\nu \hspace{0.3em} ^*F^{\mu\nu} = 0 &,& \label{4formfar}  \verb!   !
\nabla_\nu F^{\mu\nu}  =  J^\mu , \label{4formamp} 
\end{eqnarray} 
where variables with subscripts, plus/minus ($\pm$), 
are those of the fluid of positively/negatively charged particles,
$n_\pm$ is the proper particle number density, $p_\pm$ is 
the proper pressure, $h_\pm$ is the relativistic enthalpy density\footnote{The relativistic 
enthalpy includes the rest mass energy.
In the case of perfect fluid gas with specific heat ratio $\Gamma_\pm$,
it is given by $h_\pm = m_\pm n_\pm + p_\pm/(\Gamma_\pm -1) + p_\pm$. \label{entid}}, 
$U_\pm^\mu$ is the 4-velocity,
$\nabla_\mu$ is the covariant derivative,
$F_{\mu\nu}=\nabla_\mu A_\nu - \nabla_\nu A_\mu$ 
is the electromagnetic field tensor ($A_\mu$ is the 4-vector potential),
$^*F^{\mu\nu}$ is the dual tensor of $F_{\mu\nu}$,
$R^\mu$ is the frictional 4-force density between the two fluids, 
and $J^\mu$ is the 4-current density. 
We will often write a set of the spatial components of the 4-vector 
using a bold italic font, e.g., $\VEC{U}_\pm = (U^1_\pm,U^2_\pm,U^3_\pm)$, 
$\VEC{J} = (J^1,J^2,J^3)$, $\VEC{R} = (R^1, R^2, R^3)$. 
We further define the Lorentz factor $\gamma_\pm = U^0_\pm$, the 3-velocity 
$V^i_\pm=U^i_\pm/\gamma_\pm$, the electric field $E_i=F^{0i}$, the magnetic flux 
density $\sum_{k=1}^3 \epsilon^{ijk} B_k= F^{ij}$ 
($\epsilon^{ijk}$ is the Levi--Civita tensor), and the electric charge density 
$\rho_{\rm e} = J^0$. 
Here, the alphabetic index ($i,j,k$) runs from 1 to 3.

To derive one-fluid equations of the plasma, we define the average and 
difference variables as,
\begin{eqnarray} 
\rho &=& m_+ n_+ \gamma_+' + m_- n_- \gamma_-' , \label{averho} \verb!   !
n = \frac{\rho}{m} ,\\
p &=& p_+ + p_-,  \verb!   !
\Delta p = p_+ - p_-, \\
U^\mu &=& \frac{1}{\rho} ( m_+ n_+ U_+^\mu + m_- n_- U_-^\mu ) , \\
\label{ave4vel}  
J^\mu &=& e(n_+ U_+^\mu - n_- U_-^\mu) , 
\label{ave4cur}
\end{eqnarray}
where $\gamma_\pm'$ is the Lorentz factor of the two fluids observed 
by the local center-of-mass
frame of the plasma $S'$ and $m=m_+ + m_-$. Hereafter, a prime is used to denote
the variables of the center-of-mass frame.
Using these variables, we write
\begin{equation}
n_\pm U_\pm ^\mu = \frac{1}{m} \left  ( 
\rho U^\mu \pm \frac{m_\mp}{e} J^\mu \right )  .
\end{equation}
We also define the average and difference variables with respect to the
enthalpy density as
\begin{eqnarray} 
h &=& n^2 \left ( \frac{h_+}{n_+^2} + \frac{h_-}{n_-^2} \right ), 
\label{aveenth1} \\ 
\Delta h &=& \frac{n^2}{4\mu} \left ( 
\frac{h_+}{n_+^2} \frac{2m_-}{m} - \frac{h_-}{n_-^2} \frac{2m_+}{m} \right )
= \frac{mn^2}{2} \left ( \frac{h_+}{m_+ n_+^2} -  \frac{h_-}{m_- n_-^2} \right ), 
\label{avedent} \\ 
h^\ddagger &=& \frac{n^2}{4 \mu} \left [ 
\frac{h_+}{n_+^2} \left ( \frac{2m_-}{m} \right )^2 
+ \frac{h_-}{n_-^2} \left ( \frac{2m_+}{m} \right )^2 \right ], \\ 
\Delta h^\sharp &=& - \frac{n^2}{8\mu} \left [ 
\frac{h_+}{n_+^2} \left ( \frac{2m_-}{m} \right )^3 
- \frac{h_-}{n_-^2} \left ( \frac{2m_+}{m} \right )^3 \right ] , 
\end{eqnarray}
where $\mu = m_+ m_-/m^2$ is the normalized reduced mass and $\Delta \mu = (m_+ -m_-)/m$
is the normalized mass difference of the positively and negatively charged particles.
It is noted that we have a relation, $\mu = (1-\Delta \mu^2)/4$.
We find the following relations between the variables with respect to the
enthalpy density,
\begin{eqnarray}
h^\ddagger &=& h - \Delta \mu \Delta h , 
\label{aveentda} \\
\Delta h^\sharp &=& \Delta \mu h - \frac{1-3\mu}{2 \mu} \Delta h.
\end{eqnarray}

We introduce the electric resistivity $\eta$ by
\begin{equation}
\frac{R^{i'}}{ne} = - \eta J^{i'}.
\label{def_eta}
\end{equation}
According to \citet{koide08,koide09,koide10}, we introduce the scalar $\Theta$ by
\begin{equation}
\frac{R^{0'}}{n e} = \eta \rho_{\rm e}' \Theta,
\label{def_etarhoetheta}
\end{equation}
where $R^{0'}$ represents density of power (energy per unit time)
transported from the negatively
charged fluid to the positively charged fluid.
Equations (\ref{def_eta}) and (\ref{def_etarhoetheta}) yield
\begin{equation}
\frac{R^\mu}{ne} = - \eta [ J^\mu - \rho'_{\rm e} (1 + \Theta) U^\mu],
\end{equation}
where $\rho_{\rm e}'$ is the charge density observed by the local 
center-of-mass frame of the two fluids $S'$ and $\rho_{\rm e}' = - U_\nu J^\nu$.
Using the above variables, we have
one-fluid equations from the two-fluid equations 
(\ref{4formnum}) and (\ref{4formmom}),
\begin{eqnarray}
&\nabla_\nu &(\rho U^\nu) = 0 , \label{onefluidnum} \\
&\nabla_\nu & \left [ 
h U^\mu U^\nu + \frac{\mu h^\ddagger}{(ne)^2} J^\mu J^\nu 
+ \frac{\Delta h}{2ne} (U^\mu J^\nu + J^\mu U^\nu ) \right ]
= -\nabla^\mu p + J^\nu {F^\mu}_\nu , \label{onefluidmom}  \\
\frac{\mu m}{ne^2}  & \nabla_\nu & \left [  
\frac{h^\ddagger}{nm} (U^\mu J^\nu + J^\mu U^\nu ) 
+ \frac{\Delta h}{2m} e U^\mu U^\nu
- \frac{\Delta h^\sharp}{n^2 e m} J^\mu J ^\nu \right ] \nonumber \\
&=& \frac{1}{2ne} \nabla^\mu (\Delta \mu p - \Delta p) +
\left ( U^\nu - \frac{\Delta \mu}{ne} J^\nu \right) {F^\mu}_\nu 
- \eta (J^\mu - \rho'_{\rm e} (1 + \Theta) U^\mu ) .
\label{onefluidohm}
\end{eqnarray}
Equation (\ref{onefluidnum}) is the equation of continuity
and Equation (\ref{onefluidmom}) is the momentum equation.
The second and third terms in the brackets on the left-hand side
of momentum equation (\ref{onefluidmom}) are recognized as the energy-stress
tensor due to current-carrier inertia 
(e.g., electron in the case of the normal plasma).
The left-hand side of Equation (\ref{onefluidohm}) shows the inertia of
the current carrier.
With respect to the left-hand side of Equation (\ref{onefluidohm}),
we define a tensor of the electric current
\begin{equation}
q^{\mu \nu} = \frac{h^\ddagger}{n m} (U^\mu J^\nu + J^\mu U^\nu ) 
+ \frac{e \Delta h}{2m} U^\mu U^\nu - \frac{\Delta h^\sharp}{m n^2 e} J^\mu J ^\nu.
\end{equation}

Recently, we noticed that the $\Theta$ term of Equation (\ref{onefluidohm})
is determined by Equation (\ref{onefluidohm}) itself as
\begin{equation}
\eta \rho_{\rm e}' \Theta = 
- \frac{\Delta \mu}{n e} U^\rho J^\sigma F_{\sigma \rho}
+ \frac{1}{2 n e} U^\sigma \partial_\sigma (\Delta \mu p - \Delta p)
- \frac{\mu m}{ne^2}  U^\sigma \nabla_\rho q_\sigma^\rho  .
\label{etarhoetheta}
\end{equation}
The derivation of Equation (\ref{etarhoetheta}) is shown in the next subsection.
It is noted that the first term on the right-hand side of Equation (\ref{etarhoetheta}) vanishes
when the Hall effect is negligible, the second term vanishes when the
thermo-electromotive force is negligible, and the last term vanishes
when the current-carrier inertia is negligible.

%
Using Equation (\ref{4formamp}) and an equation derived by Maxwell equations
\[
(\nabla_\nu F_{\mu \sigma}) F^{\nu \sigma} = \frac{1}{4} g_{\mu\nu}
\nabla^\nu (F^{\kappa\lambda} F_{\kappa\lambda}) ,
\]
we write the equation of motion (Equation (\ref{onefluidmom})) by
\begin{equation}
\nabla_\nu T^{\mu\nu} = 0 ,
\label{eomcn}
\end{equation}
where 
\begin{equation}
T^{\mu\nu} = p g^{\mu\nu} 
+ h \left [ U^\mu U^\nu + \frac{\mu h^{\ddagger}}{(ne)^2 h} J^\mu J^\nu 
+ \frac{2 \mu \Delta h}{ne h} (U^\mu J^\nu + J^\mu U^\nu) \right ]
+ {F^\mu}_\sigma F^{\nu \sigma} - \frac{1}{4} g^{\mu\nu} 
(F^{\kappa\lambda}F_{\kappa\lambda}) .
\label{detcn}
\end{equation}
This equation corresponds to the conservation law of 4-momentum of
plasma and the electromagnetic field in the ideal GRMHD,
for example, Equation (A2) in Appendix A of \citet{koide06}.
The newly additional terms in the conservation law of 4-momentum are 
the energy-stress tensor due to current-carrier inertia, $\mu h^\ddagger J^\mu J^\nu/(ne)^2$
and $2\mu \Delta h(U^\mu J^\nu + J^\mu U^\nu)/(ne)$, in Equation (\ref{detcn}).
These additional terms express the non-MHD effects.
%
%


\subsection{Derivation of the $\Theta$ term in the generalized Ohm law}
 \label{sec3}

We derive the $\Theta$ term in the generalized Ohm law (\ref{onefluidohm}) by Equation 
(\ref{onefluidohm}) itself.
In the plasma rest frame $S'$, 
Equation (\ref{onefluidohm}) yields
\begin{eqnarray}
\frac{\mu m}{ne^2}  \nabla_{\nu'} q_{\mu'}^{\nu'}
= \frac{1}{2ne} \nabla^{\mu'} (\Delta \mu p - \Delta p) +
\left ( U^{\nu'} - \frac{\Delta \mu}{ne} J^{\nu'} \right) {F^{\mu'}}_{\nu'} 
- \eta [J^{\mu'} - \rho'_{\rm e} (1 + \Theta) U^{\mu'}].
\label{onefluidohm2uni}
\end{eqnarray}
When we take $\mu'=0$ in the equations of the generalized Ohm law, we have
\begin{eqnarray}
\frac{\mu m}{ne^2} \nabla_{\nu'} q_{0'}^{\nu'}
=  \frac{\Delta \mu}{ne} J^{i'} E_{i'}
+\frac{1}{2ne} \frac{\partial}{\partial t'} (\Delta \mu p - \Delta p)
- \eta \rho'_{\rm e} \Theta.
\label{maruc}
\end{eqnarray}
Equation (\ref{maruc}) yields Equation (\ref{etarhoetheta}) with identities,
$J^{i'} E_{i'} = J^\nu U^\mu F_{\nu \mu}$, $\displaystyle \frac{\partial}{\partial t'}
= U^\mu \partial_\mu$, and $\nabla_{\nu'} q_{0'}^{\nu'} = U^\mu \nabla_\nu q_\mu^\nu$.
The derivation of Equation (\ref{etarhoetheta}) clearly shows that
the first, second, and last terms on the right-hand side of equation (\ref{etarhoetheta})
vanish when the Hall effect, thermo-electromotive force, and 
the current-carrier inertia are negligible, respectively.
We express the form of the $\Theta$ term in several cases as follows.
\begin{itemize}
\item Standard Ohm law: 
When the Hall effect, thermo-electromotive force, and current-carrier inertia
are negligible, the $\Theta$ term vanishes 
and Equation (\ref{onefluidohm}) becomes the well-known standard relativistic Ohm law.
Then we have
\begin{equation}
\eta \rho'_{\rm e} \Theta = 0.
\end{equation}
\item A case of Hall term only:
When the electric resistivity, thermo-electromotive force, and 
current-carrier inertia are negligible, the $\Theta$ term also vanishes 
because $\VEC{E}' \cdot \VEC{J}' = \VEC{0}$, that is,
\begin{equation}
\eta \rho'_{\rm e} \Theta = 0.
\end{equation}
\item Standard Ohm law with Hall term:
When the thermo-electromotive force and current-carrier inertia
are negligible, the $\Theta$ term is given by
\begin{equation}
\eta \rho_{\rm e}' \Theta = \frac{\Delta \mu}{ne} U^\mu J^\nu F_{\nu\mu}
\end{equation}
according to Equation (\ref{etarhoetheta}).
\item A case of negligible current-carrier inertia:
When the thermo-electromotive force is significant while energy-stress tensor
of the 4-current density $q^{\mu\nu}$ is negligible, the $\Theta$ term includes 
the time-derivative of plasma pressure. Then we have
\begin{equation}
\eta \rho_{\rm e}' \Theta = \frac{\Delta \mu}{ne} U^\mu J^\nu F_{\nu\mu}
- U^\mu \partial_\mu (\Delta \mu p - \Delta p).
\end{equation}
\item Generalized Ohm law with most general form:
When the current-carrier inertia is significant, 
the energy-stress tensor of 4-current density $q^{\mu\nu}$ is not negligible and 
the $\Theta$ term becomes complex.
The term of $q^{\mu\nu}$ may be negligible 
when the Hall effect and thermo-electromotive force
are not negligible, while the terms of the Hall effect and thermo-electromotive force
are not negligible when the term of $q^{\mu\nu}$ is not negligible 
as discussed in Section \ref{sec6}.
\end{itemize}


To reveal the physical meaning
of the form of the $\Theta$ term given by Equation (\ref{etarhoetheta}), 
%
we calculated the $\Theta$ term 
in the case of iso-thermal two fluids in charge neutrality as
$p_\pm = n_0 T_0 $, $n_+ = n_- \equiv n_0$, $u_\pm^{i'} \neq 0$, $u_\pm^{i'} \neq 0$,
$\gamma_+ = \gamma_- =1$, where $T_0$ is the temperature of the two fluids.
We write $\Theta$ of this case by $\Theta_{\rm iso}$.
To keep the temperature of the two fluids equal, we have to distribute the same
amount of the thermal energy released by the Joule heating to the two fluids.
When the kinetic energy of the positively/negatively charged fluid is released
with the power density $S_+$ and $S_-$, respectively, the energy density per unit time 
transported from the negatively charged fluid to the positively charged fluid is
\begin{equation}
R^{0'} = \frac{1}{2} (S_+ + S_-) - \frac{m_-}{m_+} (S_+ + S_-)
= \frac{1}{2} \Delta \mu (S_+ + S_-).
\end{equation}
As the Joule heating is given by
$ S_+ + S_- = E_{i'} J^{i'}$,
we obtain
\begin{equation}
\eta \rho_{\rm e}' \Theta_{\rm iso} = \frac{R^{0'}}{ne} = \frac{\Delta \mu}{2 n e} E_{i'} J^{i'}
= \frac{1}{2} \frac{\Delta \mu}{ne} J^\mu U^\nu F_{\mu \nu}.
\label{etarhoe6intui}
\end{equation}
Equation (\ref{etarhoe6intui}) is a half of $\eta \rho_{\rm e}' \Theta$ in 
Equation (\ref{etarhoetheta}).
This means the Hall effect with resistivity causes the temperature difference
between the positively charged fluid and the negatively charged fluid.
In the case of a normal plasma, we find $\eta \rho_{\rm e}' \Theta > \eta \rho_{\rm e}' \Theta_{\rm iso}$ because $\Delta \mu \approx 1$, $\VEC{E}' \cdot \VEC{J}' > 0$ 
for Joule heating. Then, Joule heating causes the ion fluid temperature
to be higher than the electron temperature.

This also suggests that even in the resistive plasma the positively charged fluid
and the negatively charged fluid do not exchange their thermal energy
without the Hall effect, thermo-electromotive force, or 
current-carrier inertia effect.

\section{3+1 formalism}
 \label{sec4}

We derive a 3+1 formalism of the equations with ``the normal observer frame" 
in this paper. 
The line element of the displacement $dx^\mu$ in the spacetime 
is represented by
\begin{equation}
ds^2 =  g_{\mu \nu} dx^{\mu} dx^{\nu} 
= - \alpha^2 dt^2 + \sum_{i,j} \gamma_{ij} (dx^i +  \beta^i dt)(dx^j + \beta^j dt),
\label{defle}
\end{equation}
where
$\gamma_{ij} = g_{ij}$,
$g_{0i} = g_{ij}  \beta^j$, and 
$\alpha = \left ( - g_{00} + \sum_{i,j} g_{ij} \beta^i \beta^j \right )^{1/2} 
= \left (- g_{00} + \sum_i g_{0i} \beta^i \right )^{1/2}$.

We introduce a local inertia frame called the ``normal observer frame", 
($\tilde{t}, x^{\tilde{1}}, x^{\tilde{2}}, x^{\tilde{3}})$ as
\begin{equation}
ds^2= - d \tilde{t}^2 + \gamma_{ij} dx^{\tilde{i}} dx^{\tilde{j}}
\label{redlez}
\end{equation}
where 
\begin{eqnarray}
d\tilde{t} = \alpha dt, \verb!   !
\label{transf1}
\label{retzm}
dx^{\tilde{i}} = dx^i + \beta^i dt.
\label{transf2}
\label{rexzm}
\end{eqnarray}
Here, we have 
$g= {\rm det} (g_{\mu\nu}) = - \alpha^2 \gamma = - \alpha^2 {\rm det} (\gamma_{ij})$.
\footnote{When we write any contravariant
vector by $a^\mu$, according to Equation (\ref{transf1}), 
the contravariant vector in the normal observer frame, $a^{\tilde{\mu}}$, is given by
$a^{\tilde{0}} = \alpha a^0, 
a^{\tilde{i}} = a^i + \beta^i a^0 = a^i - \alpha N^i a^0$.
A covariant vector $a_{\tilde{\mu}}$ is 
$\displaystyle a_{\tilde{0}} = \frac{1}{\alpha} ( a_0 - \beta^i a_i)
= \frac{1}{\alpha} (a_0 +  \alpha N^i a_i) ,
a_{\tilde{i}} = a_i$.
In the normal observer frame $x^{\tilde{\mu}}$ we have
$a^{\tilde{0}} = - a_{\tilde{0}}$
and 
$a^{\tilde{i}} = \gamma^{ij} a_{\tilde{j}}$, where $\gamma_{ij} \gamma^{jk} = \delta^k_i$
($\delta_i^j$ is the Kronecker delta). \label{footnotea}}
The 4-velocity of the normal observer frame is 
$N^\mu = (1/\alpha, - \beta^i/\alpha)$,
$N_\mu = (-\alpha, 0, 0, 0)$. 
Denoting these components observed by the normal observer frame with tildes
and using the equations in footnote \ref{footnotea}, we have
\begin{eqnarray}
\gamma_{\rm L} & \equiv & U^{\tilde{0}} = \alpha U^0 , \label{deg} 
v^{\tilde{i}}  \equiv  \frac{U^{\tilde{i}}}{U^{\tilde{0}}}
=\frac{1}{\gamma} U^i - N^i \frac{U^0}{\gamma}  , \label{dev} \\
\tilde{e} & \equiv & T^{\tilde{0}\tilde{0}} = \alpha^2 T^{00} , \label{dee} 
P^{\tilde{i}}  \equiv  T^{\tilde{i}\tilde{0}} =  \alpha T^{0i} - \alpha^2 N^i T^{00} , \label{dep} \\
T^{\tilde{i}\tilde{j}} & = & T^{ij} - \alpha N^i T^{0j}
- \alpha N^j T^{i0} + \alpha^2 T^{00}  , \label{det} \\
E_{\tilde{i}} & \equiv & F_{\tilde{i}\tilde{0}} = - F_{\tilde{0}\tilde{i}} =
\frac{1}{\alpha} F_{i0} + \sum_j N^j F_{ij} , \label{del} 
\sum_k \epsilon_{ijk} B^{\tilde{k}}   \equiv  F_{\tilde{i}\tilde{j}} = F_{ij} , \label{dmg} \\ 
\tilde{\rho}_{\rm e} & \equiv & J^{\tilde{0}} = \alpha J^0 , \label{dcd} 
J^{\tilde{i}}  =   J^i - \alpha N^i J^0   . 
\label{dcu}
\end{eqnarray}

The relationship between the variables measured in the normal observer frame
is similar to that of ideal special-relativistic MHD but not identical \citep{koide96}.

The generalized GRMHD equations except for the Ohm law 
(\ref{4formfar}), (\ref{onefluidnum}), and (\ref{onefluidmom}) are written as,
\begin{eqnarray}
\frac{1}{\sqrt{-g}} \frac{\partial}{\partial x^\nu}
\left ( \sqrt{-g} \rho U^\nu \right )
= 0,
\label{eqma}
\label{eocpd}
\\
\frac{1}{\sqrt{-g}} \frac{\partial}{\partial x^\nu}
\left ( \sqrt{-g} T^{\mu \nu} \right )
+\Gamma_{\sigma \nu}^\mu T^{\sigma \nu}
= 0,
\label{eqem}
\label{eompd}
\\
\partial _\mu F_{\nu \lambda} +
\partial _\nu F_{\lambda \mu} +
\partial _\lambda F_{\mu \nu} = 0   ,
\label{eqfa}
\label{frlpd}
\\
\frac{1}{\sqrt{-g}} \frac{\partial}{\partial x^\nu}
\left ( \sqrt{-g} F^{\mu \nu} \right ) =- J^\nu   ,
\label{eqam}
\label{amlpd}
\end{eqnarray}
where we used the following relations,
$\nabla_\mu a^\nu = \partial_\mu a^\nu + \Gamma_{\mu\sigma}^\nu a^\sigma$
for any 4-vector $a^\mu$,
$\displaystyle \Gamma_{\mu\sigma}^\nu = \frac{1}{2} g^{\nu\rho} (-\partial_\rho g_{\mu\sigma}
+\partial_\mu g_{\rho\sigma} + \partial_\sigma g_{\mu\rho})$,
$\Gamma_{\mu\sigma}^\sigma = \partial_\mu (\ln \sqrt{-g})$,
and $F_{\mu\nu}=-F_{\nu\mu}$.
With respect to the Ohm law (\ref{onefluidohm}), 
we have the following form,
\begin{eqnarray}
\frac{\mu m}{ne^2} \nabla_\nu q^{\mu\nu} &=& \frac{\mu m}{ne^2} \left [ 
\frac{1}{\sqrt{-g}} \partial_\nu \left ( \sqrt{-g} q^{\mu\nu} \right )
+ \Gamma_{\sigma\nu}^\mu q^{\sigma\nu} \right ]  \nonumber \\
&=& \frac{1}{2ne} \nabla^\mu (\Delta \mu p - \Delta p)
+ \left ( U^\nu - \frac{\Delta \mu}{ne} J^\nu \right ) {F^\mu}_\nu
-\eta [J^\mu - \rho_{\rm e}'(1+\Theta) U^\mu]  .
\label{genrelgenohm4}
\label{ohm4fk}
\end{eqnarray}
Using the normal observer variables, we obtain the following set of 3+1
formalism of the general GRMHD equations
from equations (\ref{eqma})--(\ref{amlpd}) and (\ref{ohm4fk}):
\begin{eqnarray}
\frac{\partial}{\partial t} (\gamma_{\rm L} \rho) &=& - \frac{1}{\sqrt{\gamma}} \frac{\partial}{\partial x^i} \left [ \alpha \sqrt{\gamma} \gamma_{\rm L} \rho  (v^{\tilde{i}} + N^i) \right ]
\label{eqcont3+1} , \\
\frac{\partial}{\partial t} P_{\tilde{i}} & = & - \frac{1}{\sqrt{\gamma}} \frac{\partial}{\partial x^k}[ \alpha \sqrt{\gamma} (T^{\tilde{k}}_{\tilde{i}} + {N}^k P_{\tilde{i}}) ] - \frac{\partial \alpha}{\partial x^i} \tilde{e}
 - \frac{\partial}{\partial x^i} (\alpha N^k) P_{\tilde{k}} + \frac{\alpha}{2} \frac{\partial \gamma_{jk}}{\partial x^i} T^{\tilde{j}\tilde{k}}  ,
\label{3+1motot} \\
 \frac{\partial}{\partial t} \tilde{e} & = & - \frac{1}{\sqrt{\gamma}} \frac{\partial}{\partial x^i} [\alpha \sqrt{\gamma} (P^{\tilde{i}} + {N}^i \tilde{e})] - \frac{\partial \alpha}{\partial x^i} P^{\tilde{i}} 
 - \left [ \gamma_{jk} \frac{\partial}{\partial x^i} (\alpha N^k) + \frac{1}{2} \alpha N^k \frac{\partial}{\partial x^k} \gamma_{ij} \right ] T^{\tilde{i}\tilde{j}}  ,
\label{3+1entot} \\
\label{cmem}
\label{eoe31}
&& \left ( U^{\tilde{\nu}} - \frac{\Delta \mu}{ne} J^{\tilde{\nu}}  \right ) F_{\tilde{i}\tilde{\nu}}
-\eta [J^{\tilde{i}} - \tilde{\rho}_{\rm e}'(1+\Theta) U^{\tilde{i}}]  
+ \frac{1}{2ne} \frac{\partial}{\partial x^i} (\Delta \mu p - \Delta p) \nonumber \\
&=&
\frac{1}{\alpha} \frac{\mu m}{n e^2} \left [ 
\frac{\partial}{\partial t} q^{\tilde{0}}_{\tilde{i}} + \frac{1}{\sqrt{\gamma}} \frac{\partial}{\partial x^k}[ \alpha \sqrt{\gamma} (q^{\tilde{k}}_{\tilde{i}} + {N}^k q^{\tilde{0}}_{\tilde{i}}) ] +  \frac{\partial}{\partial x^i} (\alpha N^k) q^{\tilde{0}}_{\tilde{k}} - \frac{\alpha}{2} \frac{\partial \gamma_{jk}}{\partial x^i} T^{\tilde{j}\tilde{k}}
\right ] ,
\label{genrelgenohm3+1}  
\label{ohm31}   \\
\eta \rho'_{\rm e} \Theta &=& \Delta \mu J^{\tilde{\mu}} U^{\tilde{\nu}} F_{\tilde{\mu}\tilde{\nu}} + \left (
U^{\tilde{0}} \frac{1}{\alpha}  \frac{\partial}{\partial t} +
U^{\tilde{i}} \frac{\partial}{\partial x^{i}} \right ) (\Delta \mu p - \Delta p)
+ \frac{\mu m}{ne^2} U^{\tilde{\nu}} \left [
\frac{1}{\sqrt{\gamma}} 
\frac{\partial}{\partial x^{\tilde{\mu}}} ( \sqrt{\gamma} q^{\tilde{\mu}}_{\tilde{\nu}} )
- \frac{1}{2} \frac{\partial \gamma_{jk}}{\partial x^\nu} q^{\tilde{j} \tilde{k}}
\right ] ,
\label{ohm31z}   \\
\frac{\partial}{\partial t} B^{\tilde{i}} &=& - \epsilon^{{i}{j}{k}}
\frac{\partial}{\partial x^j} \left [ \alpha ( E_{\tilde{k}} - {\epsilon}_{kpq} N^p B^{\tilde{q}}) \right ] ,
\label{faraday}
\label{ampere} \\
\frac{\partial}{\partial t} E^{\tilde{i}}
&+& \alpha (J^{\tilde{i}} + \tilde{\rho}_{\rm e} N^i)
= \epsilon^{{i}{j}{k}} \frac{\partial}{\partial x^j}
[\alpha (B_{\tilde{k}} + {\epsilon}_{kmn} N^m E^{\tilde{n}})], \\
\label{ampere}
&& \frac{1}{\sqrt{\gamma}} \frac{\partial}{\partial x^i} (\sqrt{\gamma} B^{\tilde{i}}) = 0
\label{glm31} \\
&& \frac{1}{\sqrt{\gamma}} \frac{\partial}{\partial x^i} (\sqrt{\gamma} E^{\tilde{i}}) 
= \tilde{\rho}_{\rm e} .
\label{gle31} 
\end{eqnarray}
Here, we used formulae about the covariant derivative of a symmetric tensor (\ref{3+1enem}) 
and (\ref{3+1moem}) in the Appendix A.
With respect to Equation (\ref{ohm31z}), we 
assumed $\displaystyle \frac{\partial \gamma_{jk}}{\partial t}=0$ and used
$\displaystyle \nabla_{\tilde{\mu}} q^{\tilde{\mu}}_{\tilde{\nu}} 
=  \frac{1}{\sqrt{\gamma}} \frac{\partial}{\partial x^{\tilde{\mu}}} (\sqrt{\gamma} 
q^{\tilde{\mu}}_{\tilde{\nu}}) - \frac{1}{2} \frac{\partial \gamma_{jk}}{\partial x^\nu}
q^{\tilde{j} \tilde{k}}$, where
$\displaystyle \frac{\partial}{\partial x^{\tilde{0}}}=\frac{\partial}{\partial \tilde{t}} 
= \frac{1}{\alpha} \frac{\partial}{\partial t} + N^i \frac{\partial}{\partial x^i}$ and
$\displaystyle \frac{\partial}{\partial x^{\tilde{i}}} 
= \frac{\partial}{\partial x^i}$.
The terms with $\alpha$ express the gravitation and the time lapse.
The terms with $N^i$ express the frame-dragging effect around a spinning black hole.
%

\section{Characteristic parameters of relativistic plasma introduced by linear analyses 
of plasma waves}
 \label{sec5}

We newly proposed a closed \textcolor{black}{system} of generalized GRMHD equations
(\ref{onefluidnum})--(\ref{onefluidohm}) which are applicable not only for
electron-ion (normal) but also for pair
plasmas in the previous section.
In this section, we introduce the characteristic parameters of relativistic plasma
($\omega_{\rm p}$, $\omega_{\rm c}$,  $u_{\rm A}$, $c_{\rm s}$,  $\lambda_{\rm D}$,
etc.)
using the linear analyses of these equations
concerning various plasma waves and perturbations.

We investigate oscillations and waves propagating in a uniform, rest plasma
and a uniform magnetic field in the flat spacetime ($\alpha=1$ and $N^i=0$). 
For convenience, we use the 3-vector form like $\VEC{U}=(U^1,U^2,U^3)$,
$\VEC{J}=(J^1,J^2,J^3)$,$\VEC{B}=(B^1,B^2,B^3)$,$\VEC{E}=(E^1,E^2,E^3)$,
$\VEC{q}=(q^{01},q^{02},q^{03})$.
In the flat spacetime, linearized equations of perturbations,
$\tilde{\rho} =\rho - \bar{\rho}$, $\tilde{p}=p - \bar{p}$, 
$\tilde{h} = h^\dagger - \bar{h}$, $\tilde{\VEC{U}}=\VEC{U}$, 
$\tilde{\VEC{B}}=\VEC{B} - \bar{\VEC{B}}$, and $\tilde{\VEC{E}}=\VEC{E}$
are derived by Equation (\ref{eqcont3+1})--(\ref{ampere}) as
\begin{eqnarray}
\frac{\partial}{\partial t} \tilde{\rho} 
= - \bar{\rho} \nabla \cdot \tilde{\VEC{U}} ,& \displaystyle
\bar{h} \frac{\partial}{\partial t} \tilde{\VEC{U}}  = - \nabla \tilde{p}
+ \tilde{\VEC{J}} \times \bar{\VEC{B}} , \label{linrmhdnum} \\
\mu \frac{\bar{h}}{\bar{q}} \frac{\partial}{\partial t} \tilde{\VEC{q}} 
= \frac{1}{2\bar{q}} \nabla (\Delta \mu \tilde{p} - \Delta \tilde{p})
&+(\tilde{\VEC{U}} - \Delta \mu \tilde{\VEC{K}} ) \times \bar{\VEC{B}}
+\tilde{\VEC{E}} - \eta \tilde{\VEC{J}}  ,  
\label{linrmhdohm} \\
\nabla \cdot \tilde{\VEC{E}} = \tilde{\rho}_{\rm e} ,& 
\nabla \cdot \tilde{\VEC{B}} = 0  , \\
\frac{\partial}{\partial t} \tilde{\VEC{B}} = - \nabla \times \tilde{\VEC{E}} ,&
\displaystyle \tilde{\VEC{J}} + \frac{\partial}{\partial t} \tilde{\VEC{E}} 
=  \nabla \times \tilde{\VEC{B}}  \label{linrmhdamp} .
\end{eqnarray}
These equations are closed with the equation of state (EoS),
$h=h(\rho,p) = \rho H_{\rm s}(p/\rho)$.
The adiabatic EoS for single-component relativistic fluids,
which are in thermal equilibrium, has been known, and is given by
\begin{equation}
\frac{h_\pm}{\rho_\pm} = \frac{K_3(\rho_\pm/p_\pm)}{K_2(\rho_\pm/p_\pm)}
\equiv H_{\rm s} \left ( \frac{p_\pm}{\rho_\pm} \right )
\label{chandragynge}
\end{equation}
\citep{chandrasekhar38,synge57}. Here, $K_2$ and $K_3$ are the
modified Bessel functions of the second kind of order two and three,
respectively. 
When we consider the adiabatic one-component fluid
in the rest frame, 
the adiabatic condition\footnote{When we consider a fluid element of particle number $N$,
volume $V$, and enthalpy $H$, the first law of thermodynamics is
$dH = d'Q + V dp$, where $d'Q$ is the heat energy from the outside of the
fluid and vanishes in the adiabatic case. Using $H=hV$, $N=n V$,
$d'Q=0$, we easily obtain Equation (\ref{tilden}).} yields,
\begin{equation}
\frac{\tilde{n}_\pm}{\bar{n}_\pm} 
= \frac{\tilde{h}_\pm - \tilde{p}_\pm}{\bar{h}_\pm} .
\label{tilden}
\end{equation}
From equation (\ref{chandragynge}), we find
\begin{equation}
\frac{\bar{\rho}_\pm}{\bar{p}_\pm} \frac{\tilde{p}_\pm}{\tilde{\rho}_\pm}
=\frac{H_{\rm s}' \left (\frac{\bar{p}_\pm}{\bar{\rho}_\pm} \right )}
{H_{\rm s}' \left (\frac{\bar{p}_\pm}{\bar{\rho}_\pm} \right) -1}
\equiv \Gamma_{\rm a}  \left (\frac{\bar{p}_\pm}{\bar{\rho}_\pm} \right) .
\label{polytropicindex}
\end{equation}
In general, $\Gamma_{\rm a}$ is not constant and is a function of 
$\bar{p}_\pm/\bar{\rho}_\pm$. 
$\Gamma_{\rm a}$ is called the effective adiabatic index.\footnote{The polytropic
index is given by $N_{\rm pol}=(\Gamma_{\rm a} - 1)^{-1}$.}
Using $\Gamma_{\rm a}$, we obtain the EoS for the plasma
\citep{koide10} as,
\begin{eqnarray}
h &=& n \left [  H_{\rm s} \left ( \frac{p+\Delta p}{2 \rho_+} \right ) \frac{m_+^2}{\rho_+} 
+  H_{\rm s} \left ( \frac{p - \Delta p}{2 \rho_-} \right ) \frac{m_-^2}{\rho_-} \right ]  ,
\label{eoshh} \\
\Delta h &=& 2 n^2 \mu m \left [  H_{\rm s} \left ( \frac{p+\Delta p}{2\rho_+} \right ) 
\frac{m_+}{\rho_+} 
-  H_{\rm s} \left ( \frac{p-\Delta p}{2 \rho_-} \right ) \frac{m_-}{\rho_-} \right ] ,
\label{eosdh}
\end{eqnarray}
where 
\begin{equation}
\rho_\pm \equiv \left [ \rho^2 \mp \frac{m_\mp \rho}{e} U^\nu J_\nu 
-\left ( \frac{m_\mp}{e} \right )^2 J^\nu J_\nu \right ]^{1/2}.
\end{equation}
When $\bar{p}_\pm \ll \bar{\rho}_\pm $ or $\bar{p}_\pm \gg \bar{\rho}_\pm $,
$\Gamma_{\rm a}$ is asymptotically constant 
($\Gamma_{\rm a}(\bar{p}_+/\bar{\rho}_+) \approx \Gamma_{\rm a}(\bar{p}_-/\bar{\rho}_-)$),
thus we have $(1/\bar{T}) \tilde{p} = \Gamma_{\rm a} \tilde{\rho}$,
that is, 
\begin{equation}
\frac{\bar{\rho}}{\bar{p}} \frac{\tilde{p}}{\tilde{\rho}} 
= \frac{\bar{\rho}}{\bar{p}} \left ( \overline{\frac{d p}{d \rho}} 
\right )
=  \overline{ \left ( \frac{d \ln p}{d \ln \rho} \right )}
= \Gamma_{\rm a} \left ( \frac{\bar{p}}{\bar{\rho}} \right ).
\end{equation}
In general, 
$\displaystyle \frac{\bar{\rho}}{\bar{p}} \frac{\tilde{p}}{\tilde{\rho}} 
= \Gamma_{\rm a} (\bar{p},\bar{\rho})$
depends on both $\bar{\rho}$ and $\bar{p}$.

\subsection{Longitudinal modes of plasma waves and oscillations}

First, we derive a dispersion relation of longitudinal oscillation 
modes ($\tilde{\VEC{U}} \parallel \VEC{k}, 
\tilde{\VEC{J}} \parallel \VEC{k}$) in an unmagnetized, rest plasma
with uniform, finite pressure $\bar{p}$. For simplicity, we assume the
temperatures of the two fluids are the same: $\bar{T}=\bar{T}_+=\bar{T}_-$.
Using Equation (\ref{averho}) and the zeroth component of
Equation (\ref{ave4cur}), we have
$\gamma n_\pm = (\gamma \rho \pm m_\mp \rho_{\rm e}/e)/m$ when
$\gamma = \gamma_+ \approx \gamma_-$. Using these equations, we have
\begin{equation}
\frac{\partial \gamma n}{\partial t} = - \frac{2}{m} \nabla \cdot (\rho \VEC{U})
+ \frac{\mu}{e} (\nabla \cdot \VEC{J}).
\end{equation}
In the present non-relativistic case, $\gamma=1$, we have
\begin{equation}
\frac{\partial n}{\partial t} = - \frac{2}{m} \nabla \cdot (\rho \VEC{U})
+ \frac{\mu}{e} (\nabla \cdot \VEC{J}).
\end{equation}
If $\Gamma_{\rm a}(\bar{p}_+/\bar{\rho}_+) = \Gamma_{\rm a}(\bar{p}_-/\bar{\rho}_-) 
\equiv \Gamma_{\rm a}$ is uniform and constant, we obtain
\begin{eqnarray}
\frac{\partial \tilde{p}}{\partial t} &=& 
- 2 \Gamma_{\rm a} \bar{T} \left [\frac{1}{m} \nabla \cdot (\bar{\rho} \tilde{\VEC{U}}) 
- \frac{\Delta \mu}{2e} (\nabla \cdot \tilde{\VEC{J}}) \right ], \\
\frac{\partial}{\partial t} \Delta \tilde{p} &=& 
- \frac{\Gamma_{\rm a} \bar{T}}{e} \nabla \cdot \tilde{\VEC{J}} .
\end{eqnarray}
We have the linearized equations,
\begin{eqnarray}
\bar{h} \frac{\partial}{\partial t} \tilde{\VEC{U}} &=& - \nabla \tilde{p}, \\
\frac{\mu \bar{h}}{(\bar{n} e)^2} \frac{\partial}{\partial t} \tilde{\VEC{J}} &=& 
\frac{1}{2\bar{n} e} \nabla (\Delta \mu \tilde{p} - \Delta \tilde{p} ) 
+ \tilde{\VEC{E}}, \\
\tilde{\VEC{J}} + \frac{\partial}{\partial t} \tilde{\VEC{E}} &=& \VEC{0}.
\end{eqnarray}
These equations yield
\begin{eqnarray}
\frac{\mu \bar{h}}{(\bar{n}e)^2} \frac{\partial^2}{\partial t^2} \tilde{\VEC{J}} &=& 
-\frac{\Delta \mu \Gamma_{\rm a} \bar{T}}{e} \nabla (\nabla \cdot \tilde{\VEC{U}})
+\frac{\Gamma_{\rm a} \bar{T}}{2\bar{n} e^2} (1 + (\Delta \mu)^2) 
\nabla (\nabla \cdot \tilde{\VEC{J}}) - \tilde{\VEC{J}}, 
\label{oscill2j} \\
\bar{h} \frac{\partial^2}{\partial t^2} \tilde{\VEC{U}} &=& 
\frac{2 \Gamma_{\rm a} \bar{T} \bar{\rho}}{m} \nabla (\nabla \cdot \tilde{\VEC{U}}) 
-\frac{\Delta \mu \Gamma_{\rm a} \bar{T}}{e} \nabla (\nabla \cdot \tilde{\VEC{J}}) .
\label{oscill2u}
\end{eqnarray}
When $\tilde{\VEC{U}}$ vanishes, Equation (\ref{oscill2j}) yields
\begin{equation}
\frac{\mu \bar{h}}{(\bar{n} e)^2} \omega^2 \tilde{\VEC{J}} =
\frac{\Gamma_{\rm a} \bar{T}}{2\bar{n} e^2} (1 + (\Delta \mu)^2) 
\VEC{k} (\VEC{k} \cdot \tilde{\VEC{J}}) - \tilde{\VEC{J}}, 
\end{equation}
Using the condition $\tilde{\VEC{J}} \parallel \VEC{k}$ in the longitudinal mode,
we have the dispersion relation,
\begin{equation}
\omega^2 = \frac{1- 2 \mu}{\mu} c_{\rm s}^2 k^2 + \omega_{\rm p}^2,
\label{dispplasmaosci}
\end{equation}
where $\displaystyle c_{\rm s} = \sqrt{\frac{\Gamma_{\rm a} p}{h}} 
= \sqrt{\frac{dp}{d\rho} \frac{\rho}{h}}$ is the sound speed
and $\displaystyle \omega_{\rm p} = \sqrt{\frac{(ne)^2}{\mu h}}$. 
Note that $\omega_{\rm p}$ is related to the electron/ion plasma frequency
and then we call $\omega_{\rm p}$ the ``{\it unified} plasma frequency".
When we take $c_{\rm s} = c_{\rm s}^0 = \sqrt{p/h}$, 
$\omega=0$ and $k=i \sqrt{2}/\lambda_{\rm D}$, we have the ``{\it extended} Debye length",
\begin{equation}
\lambda_{\rm D} = \sqrt{\frac{2(1-2 \mu)}{\mu}}
\frac{c_{\rm s}^0}{\omega_{\rm p}}
=\sqrt{2(1-2 \mu)} \frac{1}{\omega_{\rm p}} \sqrt{\frac{p}{\mu h}} ,
\end{equation}
which expresses the characteristic length of shielding
of the electric field around an electric charge.
Using the extended Debye length, the dispersion relation (\ref{dispplasmaosci}) is written by
\begin{equation}
\omega^2 = \omega_{\rm p}^2 \left ( 1 + \frac{\Gamma_{\rm a}}{2} \lambda_{\rm D}^2 k^2
\right ).
\end{equation}

With respect to the plasma oscillation mode, 
because $\tilde{\VEC{J}} \parallel \VEC{k}$ and 
$\tilde{\VEC{U}} \parallel \VEC{k}$ in the longitudinal modes, 
Equations (\ref{oscill2j}) and (\ref{oscill2u}) yield
\begin{eqnarray}
-\omega^2 \frac{\mu \bar{h}}{(\bar{n} e)^2} \tilde{J}_\parallel &=&
\frac{\Delta \mu \Gamma_{\rm a} \bar{T}}{e} k^2 \tilde{U}_\parallel
-\frac{\Gamma_{\rm a} \bar{T}}{2\bar{n}e^2} (1+\Delta \mu^2) k^2 \tilde{J}_\parallel
- \tilde{J}_\parallel, \\
-\omega^2 \bar{h} \tilde{U}_\parallel &=&
- \frac{2 \Gamma_{\rm a} \bar{T} \bar{\rho}}{m} k^2 \tilde{U}_\parallel
+\frac{\Delta \mu \Gamma_{\rm a} \bar{T}}{e} k^2 \tilde{J}_\parallel ,
\end{eqnarray}
where $\tilde{J}_\parallel \equiv (\VEC{J} \cdot \VEC{k})/k$ 
and $\tilde{U}_\parallel \equiv (\VEC{U} \cdot \VEC{k})/k$, $k \ne 0$.
Then, we get the following dispersion relation,
\begin{equation}
\left [ 
\omega^2 - \frac{c_{\rm s}^2}{2\mu} (1+\Delta \mu^2) k^2 - \omega_{\rm p}^2
\right ]
(\omega^2 - 2 c_{\rm s}^2 k^2) =
\frac{\Delta \mu^2}{\mu} (c_{\rm s} k)^4   .
\end{equation}

In the case of an electron-ion (normal) plasma 
($\Delta \mu \approx 1$, $\mu = m_{\rm e}/m \ll 1$),
$\omega \gg c_{\rm s} k/\sqrt{\mu}$, 
the dispersion relation becomes
\begin{equation}
\omega^2 = \frac{c_{\rm s}^2}{2 \mu} k^2 + \omega_{\rm p}^2.
\end{equation}
This expression shows the dispersion relation of the plasma oscillation
for the plasma with finite pressure.
In the case of $\omega^2 \ll \omega_{\rm p}^2/\mu$, we have the dispersion relation
of sound waves
\begin{equation}
\omega^2 = 2 c_{\rm s}^2 k^2 .
\end{equation}
In the case of a pair plasma ($\Delta \mu = 0$, $\mu = 1/4$), we have
two modes
\begin{equation}
\omega^2 =  \omega_{\rm p}^2 +  2 c_{\rm s}^2 k^2 ,
\end{equation}
and
\begin{equation}
\omega^2 =  2 c_{\rm s}^2 k^2  .
\end{equation}
The former is the dispersion relation of plasma oscillation and
the latter is that of the sound wave.

Next, we investigate the bulk compressional wave of the magnetized plasma.
When $\VEC{k} \parallel \bar{\VEC{B}}$ and 
$\tilde{\VEC{U}} \parallel \bar{\VEC{B}}$, 
the dispersion relation is the
same as that of the non-magnetized plasma wave. Then, we investigate 
the case of
$\VEC{k} \perp \bar{\VEC{B}}$, $\VEC{k} \perp \tilde{\VEC{B}}$, 
$\VEC{k} \parallel \tilde{\VEC{U}}$,
and $\rho_{\rm e} = 0$. Because $\rho_{\rm e}=0$, we have
$\Delta p = 0$. In the case of 
$k \ll \omega_{\rm p}$ and $\omega \ll \omega_{\rm c}$, 
the left-hand side, the first term and the Hall term
on the right-hand side of Ohm law (\ref{linrmhdohm}) are negligible.
Using the linearized equations (\ref{linrmhdnum})--(\ref{linrmhdamp}), 
we have the dispersion relation,
\begin{equation}
\omega^2 = v_{\rm f}^2 k^2 , \label{disprelfast}
\end{equation}
where $\displaystyle v_{\rm f}^2 = \frac{\Gamma_{\rm a} \bar{p} + \bar{B}^2}{\bar{h} + \bar{B}^2}$
is the 3-velocity of the fast wave.
It is also noted that $v_{\rm f} < 1$.

\subsection{Transverse wave propagating along the magnetic field}

We investigate transverse waves propagating through the ideal
MHD plasma along the magnetic field lines,
\begin{equation}
\bar{\VEC{B}} \parallel \VEC{k}, \hspace{1cm}
\tilde{\VEC{E}}, \verb! ! \tilde{\VEC{B}}, \verb! ! \tilde{\VEC{U}} 
\parallel \VEC{k}, \hspace{1cm} \eta = 0.
\end{equation}
We assume that any perturbation $\tilde{A}$ is proportional to 
$\exp(i \VEC{k} \cdot \VEC{r} - i \omega t)$. 
The linearized equations become
\begin{eqnarray}
- i \omega \bar{h} \tilde{\VEC{U}} &=& 
\tilde{\VEC{J}} \times \bar{\VEC{B}} ,\\
- i \omega \frac{\mu \bar{h}}{\bar{n}e} \tilde{\VEC{J}}
&=& (\bar{n}e \tilde{\VEC{U}} - \Delta \mu \tilde{\VEC{J}}) \times \bar{\VEC{B}}
+ \bar{n}e \tilde{\VEC{E}}  , \\
i \VEC{k} \cdot \tilde{\VEC{E}} &=& 0 , \verb!   !
i \VEC{k} \cdot \tilde{\VEC{B}} = 0 ,\\
- i \omega \tilde{\VEC{B}} &=& 
- i \VEC{k} \times \bar{\VEC{E}} , \verb!   !
\tilde{\VEC{J}} - i \omega \tilde{\VEC{E}} = 
 i \VEC{k} \times \bar{\VEC{B}} .
\end{eqnarray}
From these linearized equations, we have
\begin{equation}
- \omega^2 \frac{\mu \bar{h}}{\bar{n}e}  \Box \tilde{\VEC{J}}
= - \frac{\bar{n}e}{\bar{h}} (\Box \tilde{\VEC{J}}) \bar{B}^2 
+ i \omega \Delta \mu  (\Box \tilde{\VEC{J}}) \times \bar{\VEC{B}}
- \omega^2 \bar{n}e \tilde{\VEC{J}}  , 
\label{lineq4transmod}
\end{equation}
where $\Box = \partial_\nu \partial^\nu = \omega^2 - k^2$
is the d'Alembertian. 

When we set the unit basis vector of the $z$-direction as $\VEC{k}=k \VEC{e}_z$,
we have $\VEC{B}_0=B_0 \VEC{e}_z$ in the present case.
Here, we make the complex function $\tilde{\mathscr{J}} \equiv 
\tilde{J}_x - i \tilde{J}_y$ which corresponds 
to the perturbation of the current density $\tilde{\VEC{J}}$. 
Equation (\ref{lineq4transmod}) is written as 
\begin{equation}
- \frac{\mu \bar{h}}{\bar{n}e} \omega^2 (\omega^2 - k^2) \tilde{\mathscr{J}}
= - \frac{\bar{n}e}{\bar{h}} (\omega^2 - k^2) \tilde{\mathscr{J}} \bar{B_0}^2 
+ \Delta \mu  \omega^2 (\omega^2 - k^2) ( B_0 \tilde{\mathscr{J}}) 
- \bar{n}e \omega^2 \tilde{\mathscr{J}} .
\end{equation}
Then, we obtain the dispersion relation of the transverse modes,
\begin{equation}
(\omega^2 - k^2) \left ( \frac{\mu h}{n^2e^2} \omega^2 - \frac{\Delta \mu B_0}{ne} \omega
- \frac{B_0^2}{h} \right ) - \omega^2 =0,
\end{equation}
that is,
\begin{equation}
(\omega^2 - k^2) \left ( \frac{1}{\omega_{\rm p}^2} \omega^2 
- \frac{\Delta \mu \omega_{\rm c}}{\omega_{\rm p}^2} \omega
- u_{\rm A}^2 \right ) - \omega^2 =0,
\label{disprel2transmod}
\end{equation}
where 
$\displaystyle u_{\rm A}=\sqrt{\frac{B_0^2}{h}}$ is the 4-Alfven velocity and 
$\displaystyle \omega_{\rm c} = \frac{eB_0}{\mu m} \frac{\rho}{h}$.
Note that $\omega_{\rm c}$ corresponds
to the cyclotron frequency, which we call the ``{\it extended} cyclotron frequency".
If we set the pressure to be zero, $\omega_{\rm c}$ reduces to the cyclotron 
frequency of the charged particle with mass $m$ and charge $e$
in the magnetic field ${B}$, $\omega_{\rm c}=e{B}/m$.
In general, we have the relation between the plasma parameters,
\begin{equation}
\frac{\omega_{\rm c}}{\omega_{\rm p}} = \frac{u_{\rm A}}{\sqrt{\mu}}.
\end{equation}
When we consider the limit $\omega \gg \omega_{\rm p}, \omega_{\rm c}$,
the dispersion relation (\ref{disprel2transmod}) yields
\begin{equation}
\omega = \pm \sqrt{k^2 + \omega_{\rm p}^2} + \frac{\Omega_{\rm FR}}{2} = \omega_{\pm},
\end{equation}
where
\begin{equation}
\Omega_{\rm FR} = \omega_+ + \omega_- 
= \frac{\Delta \mu \omega_{\rm c} \omega_{\rm p}^2}
{k^2 + \omega_{\rm p}^2 - (\Delta \mu \omega_{\rm c})^2}
\approx \frac{\Delta \mu \omega_{\rm c} \omega_{\rm p}^2}{k^2}
\end{equation}
presents the angular velocity of electromagnetic wave polarity of Faraday rotation.

\section{Estimation of non-ideal MHD terms of generalized GRMHD equations}
\label{sec6}

Here, we summarize a complete \textcolor{black}{system} of the generalized GRMHD equations 
(Equations (\ref{onefluidnum})--(\ref{onefluidohm}), and (\ref{etarhoetheta}))
derived from the general-relativistic two-fluid equations as
\begin{eqnarray}
&\nabla_\nu &(\rho U^\nu) = 0 , \label{mdfgrmnum} \\
&\nabla_\nu & \left [ 
h U^\mu U^\nu + \frac{\mu h^\ddagger}{(ne)^2} J^\mu J^\nu 
+ \frac{\Delta h}{2ne} (U^\mu J^\nu + J^\mu U^\nu ) \right ]
= -\nabla^\mu p + J^\nu {F^\mu}_\nu , \label{mdfgrmmom}  \\
U^\nu {F^\mu}_\nu &=& \eta [ J^\mu - \rho_{\rm e}' (1 + \Theta)  U^\mu ] +
\frac{\Delta \mu}{ne} J^\nu {F^\mu}_\nu 
\label{mdfgrmohm}
- \frac{1}{2ne} \nabla^\mu (\Delta \mu p - \Delta p) 
+ \frac{\mu m}{ne^2} \nabla_\nu q^{\mu \nu} , \\
\eta \rho_{\rm e}' \Theta &=& 
- \frac{\Delta \mu}{n e} U^\rho J^\sigma F_{\sigma \rho}
+ \frac{1}{2 n e} U^\sigma \partial_\sigma (\Delta \mu p - \Delta p)
- \frac{\mu m}{ne^2}  U^\sigma \nabla_\rho q_\sigma^\rho  , 
\label{etarhoetheta2}
\end{eqnarray}
where 
$\displaystyle q^{\mu\nu} = \frac{h^\ddagger}{n m} (U^\mu J^\nu + J^\mu U^\nu)
- \frac{\Delta h^\sharp}{n^2 e m} J^\mu J^\nu + \frac{2 \Delta h}{m} e U^\mu U^\nu$
is a tensor of the electric current.
It is noted that Equations (\ref{mdfgrmohm}) and (\ref{etarhoetheta2}) 
are not independent because the latter comes from the former.

We evaluate the significance of the non-ideal MHD terms in the generalized GRMHD equations
in plasma.
We introduce the ``primary (primitive)" parameters of the plasmas as follows.
In the SI units, the primary plasma parameters are written as
\begin{equation}
\omega_{\rm p}^{\rm prm} = \sqrt{\frac{n e^2}{\mu m \epsilon_0}} , \verb!   !
\omega_{\rm c}^{\rm prm} = \frac{eB}{\mu m} , \verb!   !
v_{\rm A}^{\rm prm} = \sqrt{\frac{B^2}{\mu_0 \rho}} , \verb!   !
c_{\rm s}^{\rm prm} = \sqrt{\Gamma_{\rm a} \frac{p}{\rho}},
\end{equation}
where $\epsilon_0$ and $\mu_0$ are the permittivity and permeability of vacuum, respectively.
Using the primary plasma parameters, we write the plasma parameters
introduced in this paper as
\begin{equation}
\omega_{\rm p} = \omega_{\rm p}^{\rm prm} \sqrt{f_{\rm rel}}, \verb!   !
\omega_{\rm c} = \omega_{\rm c}^{\rm prm} f_{\rm rel}, \verb!   !
u_{\rm A} = v_{\rm A}^{\rm prm} \sqrt{f_{\rm rel}}, \verb!   !
c_{\rm s} = c_{\rm s}^{\rm prm} \sqrt{f_{\rm rel}},
\end{equation}
where $f_{\rm rel} = \rho/h \le 1$ is the relativistic factor of the internal 
energy of the plasma.
The primary plasma parameters are calculated as 
\begin{eqnarray}
\omega_{\rm p}^{\rm prm} &=& 5.64 \times 10^{6}
\left ( \frac{n}{10^4 {\rm cm^{-3}}} \right )^{1/2}
\left ( \frac{\mu m}{m_{\rm e}} \right )^{-1/2} {\rm s^{-1}}, \\
\omega_{\rm c}^{\rm prm} &=& 1.76 \times 10^{7}
\left ( \frac{B}{1 {\rm G}} \right )
\left ( \frac{\mu m}{m_{\rm e}} \right )^{-1} {\rm s^{-1}}, \\
v_{\rm A}^{\rm prm} &=& 2.18 \times 10^{9}
\left ( \frac{B}{1 {\rm G}} \right )
\left ( \frac{n}{10^4 {\rm cm^{-3}}} \right )^{-1/2}
\left ( \frac{m}{m_{\rm i}} \right )^{-1} {\rm cm \, s^{-1}}, \\
c_{\rm s}^{\rm prm} &=& 1.05 \times 10^{9}
\left ( \frac{k_{\rm B} T}{10^{10} {\rm K}} \right )^{1/2}
\left ( \frac{m}{m_{\rm i}} \right )^{-1/2} 
\left ( \frac{\Gamma_{\rm a}}{4/3} \right )^{1/2} {\rm cm \, s^{-1}} .
\end{eqnarray}

Using the plasma parameters, we estimate the significance of the non-ideal MHD terms in the generalized
GRMHD equations: the terms of resistive electromotive force, Hall effect,
thermo-electromotive force, and current-carrier inertia
in the generalized Ohm law (\ref{mdfgrmohm})
and the energy-stress of electric 4-current density in Equation (\ref{mdfgrmmom}).
The left-hand side of Equation (\ref{mdfgrmohm}) is written as 
$U^\nu F_{\mu i} = \gamma E_i + \epsilon_{ijk} U^j B^k$.
We compare the terms on the right-hand side of Equation (\ref{mdfgrmohm}) 
with the term, $\epsilon_{ijk} U^j B^k$.
We take the 4-Alfven velocity $u_{\rm A}$ as the characteristic value of 4-velocity
of the plasma, $U = \sqrt{U^i U_i}$.

In this paper, we consider both the the normal plasma (ion-electron plasma) and the pair
plasma (positron-electron plasma). In the case of the normal plasma, we have
$\mu m = m_{\rm e}$ and $n=\rho/m \approx n_{\rm i} \gamma_{\rm i}' 
\approx n_{\rm i} \approx n_{\rm e}$. In the pair plasma case, we have
$\mu m = m_{\rm e}/2$ and $n=\rho/m \approx (1/2) (n_{\rm e^+} \gamma_{\rm e^+}' 
+ n_{\rm e^-} \gamma_{\rm e^-}') \approx  n_{\rm e}$.
Here, we use the condition of $m_{\rm i}/m_{\rm e} = 1836 \gg 1$,
the charge neutrality ($n_+ \approx n_-$), and $\gamma_+' \approx \gamma_-' \approx 1$.
In both cases, we approximately have $n \approx n_{\rm e}$.
Then, we have 
\begin{equation}
\eta = \frac{m_{\rm e} \nu_{+-}}{2 n_{\rm e} e^2}
= \frac{\mu m \nu_{\rm ei}}{2 n e^2} \frac{\nu_{+-}}{\nu_{\rm ei}} \frac{m_{\rm e}}{\mu m}
= \frac{\mu_0 c^2 \nu_{\rm ei}}{2 (\omega_{\rm p}^{\rm prm})^2} 
\frac{\nu_{+-}}{\nu_{\rm ei}} \frac{m_{\rm e}}{\mu m}
\approx \frac{\mu_0 c^2 \nu_{\rm ie}}{2 \omega_{\rm p}^2}
\frac{\nu_{+-}}{\nu_{\rm ei}} \frac{m_{\rm e}}{\mu m},
\end{equation}
where $\nu_{+-}$ is the collision frequency between the + and $-$ particles and
$\nu_{\rm ei}$ is the collision rate between the electrons and ions \citep{miyamoto87,miyamoto89},
\begin{equation}
\nu_{\rm ei} = \frac{n_{\rm i} e^4 \ln \Lambda}{25.8 \pi^{1/2} \epsilon_0^2
m_{\rm e}^{1/2} T_{\rm e}^{3/2}}
= 8.3 \times 10^{-13} \left ( \frac{T_{\rm e}}{10^{10} {\rm K}} \right )^{-3/2}
\left ( \frac{n_{\rm e}}{10^{4} {\rm cm^{-3}}} \right ) [{\rm s^{-1}}],
\end{equation}
where $\ln \Lambda$ is the Coulomb logarithm ($\ln \Lambda \approx 20$).

We evaluate the significance of the non-ideal MHD terms of the
generalized relativistic Ohm law as follows.
Here, we use $J \sim B/(\mu_0 L)$, $U \sim u_{\rm A}$, $\tau = L/U$,
and the relation,
\begin{equation}
\frac{\omega_{\rm c}}{\omega_{\rm p}} = \frac{u_{\rm A}}{\sqrt{\mu}} .
\label{relomcomp2ua}
\end{equation}
\begin{itemize}
\item The electric resistivity term:
\begin{equation}
\mathcal{I}_{\rm r} = \frac{\eta J}{U B} = \frac{\eta}{\mu_0 U L} = \frac{1}{S_{\rm M}}
= \frac{\eta}{\mu_0 u_{\rm A} L}
= \frac{1}{2}  \frac{\mu m}{m_{\rm e}}
\frac{\nu_{\rm ei}}{\omega_{\rm c}} \frac{\nu_{+-}}{\nu_{\rm ei}}
\frac{1}{\mu \omega_{\rm c} \tau}
= f_{\rm r} \frac{1}{\mu \omega_{\rm c} \tau},
\end{equation}
where $\displaystyle  f_{\rm r} = \frac{1}{2}  \frac{\mu m}{m_{\rm e}}
\frac{\nu_{+-}}{\omega_{\rm c}}$.
Here, in the normal or pair plasma, we have $\displaystyle \frac{\mu m}{m_{\rm e}} < 1$.
Furthermore, in a thin plasma, like a plasma around a super-massive black hole, we have
$\frac{\nu_{+-}}{\omega_{\rm c}} < 1$ (usually $\frac{\nu_{+-}}{\omega_{\rm c}} \ll 1$). 
Then, we usually use $f_{\rm r} < 1$.
\item The Hall term:
\begin{equation}
\mathcal{I}_{\rm H} = \frac{\frac{\Delta \mu}{ne} JB}{UB}
= \frac{\Delta \mu B/\mu_0 L}{n e u_{\rm A}}
= \frac{\Delta \mu u_{\rm A}}{L \mu \omega_{\rm c}}
= \Delta \mu \frac{1}{\mu \omega_{\rm c} \tau} 
= f_{\rm H} \frac{1}{\mu \omega_{\rm c} \tau} ,
\end{equation}
where $f_{\rm H} = \Delta \mu$.
In the normal plasma, we have $\Delta \mu = 1 - 2 m_{\rm e}/2m_{\rm i} < 1$.
In the pair plasma, we have $\Delta \mu = 0$.
We also have $f_{\rm H} < 1$.
\item
Thermo-electromotive force term:
\begin{equation}
\mathcal{I}_{\rm th} = \frac{\frac{1}{ne} \frac{p}{L}}{UB}
= \frac{\beta_{\rm p} B^2/2 \mu_0}{n e u_{\rm A} B L}
= \frac{\beta_{\rm p}}{2} \frac{1}{\mu \omega_{\rm c} \tau}
= f_{\rm th}  \frac{1}{\mu \omega_{\rm c} \tau} ,
\end{equation}
where $\beta_{\rm p} = p/(B^2/2 \mu_0)$ is the plasma beta
and $f_{\rm th} = \beta_{\rm p}/2$.
In the magnetically dominated plasma, we have $\beta_{\rm p} \lesssim 2$.
Then, we also have $f_{\rm th} \lesssim 1$.
\item 
Current-carrier inertia term:
\begin{equation}
\mathcal{I}_{\rm cci} = \frac{\frac{\mu h^{\ddagger}}{(ne)^2} \frac{JU}{L}}{UB}
= \frac{\mu h^{\ddagger}}{(ne)^2 L^2 \mu_0}
= \mu \frac{h^{\ddagger}}{h} \frac{1}{(\mu \omega_{\rm c} \tau)^2} 
= \frac{h^{\ddagger}}{h} \frac{1}{(\sqrt{\mu} \omega_{\rm c} \tau)^2} 
= \left ( f_{\rm cci} \frac{1}{\sqrt{\mu} \omega_{\rm c} \tau} \right )^2 ,
\end{equation}
where $f_{\rm cci} = \sqrt{h^{\ddagger}/h} = \sqrt{1 - \Delta \mu \Delta h/h}$.
When the plasma temperature is not relativistic, we have $f_{\rm cci} \sim 1$.
However, when the plasma temperature is relativistic, $f_{\rm cci}$
may become $\sim 1/\sqrt{\mu}$, maximumly.
\end{itemize}



We also evaluate the significance of the non-ideal MHD term of
the momentum equation (\ref{mdfgrmmom}).
The non-ideal MHD term of Equation (\ref{mdfgrmmom}) is the second and third terms
in the brackets on the left-hand side of Equation (\ref{mdfgrmmom}),
which is the energy-stress tensor due to the current-carrier inertia.
\begin{itemize}
\item Energy-stress tensor due to the current-carrier inertia on the momentum equation 
(\ref{mdfgrmmom}):
\begin{equation}
\mathcal{I}^{\rm meq}_{\rm cci} = \frac{\frac{\mu h^{\ddagger}}{(ne)^2} \frac{JJ}{L}}{UU}
= \mu \frac{h^{\ddagger}}{h} \left ( \frac{B/\mu_0 L}{n e u_{\rm A}} \right )^2
= \mu \frac{h^{\ddagger}}{h} \left ( \frac{B}{\omega_{\rm p}^2 u_{\rm A} L} \right )^2
= \frac{h^{\ddagger}}{h} \left ( \frac{c}{\omega_{\rm p} L} \right )^2
= \left ( \sqrt{\frac{h^\ddagger}{h}} \frac{c}{\omega_{\rm p} L} \right ) ^2
= \left ( f_{\rm cci} \frac{c}{\omega_{\rm p} L} \right ) ^2 .
\end{equation}
When we use $U = L/\tau \sim u_{\rm A}$ and Equation (\ref{relomcomp2ua}), we have
\begin{equation}
\mathcal{I}^{\rm meq}_{\rm cci} = 
\left (f_{\rm cci} \frac{c}{\omega_{\rm p} u_{\rm A} \tau} \right )^2
= \left (f_{\rm cci} \frac{1}{\sqrt{\mu} \omega_{\rm c} \tau} \right )^2
= \mathcal{I}^{\rm Ohm}_{\rm cci} .
\end{equation}
We found the non-ideal MHD conditions of the current-carrier inertia
term of the generalized Ohm law and the generalized momentum equation are identical.
Then, we use the characteristic scale of the non-ideal MHD effect due to
current-carrier inertia of the generalized Ohm law.
\end{itemize}

According to the above estimation, we conclude that the non-ideal MHD effects
are the resistive electromotive force , Hall effect, thermo-electromotive force, 
and current-carrier inertia. With respect to the three former effects, 
putting aside the details of the factors $f_{\rm r}, f_{\rm H}, f_{\rm th} \lesssim 1$, 
we have the characteristic time scale 
\footnote{The more precise characteristic time scale of resistive, Hall, 
thermo-electromotive force, and current-carrier inertia effects are given by 
$\tau_{\rm r} = f_{\rm r} \tau^{\rm Ohm}_{\rm c}$,
$\tau_{\rm H} = f_{\rm H} \tau^{\rm Ohm}_{\rm c}$, 
$\tau_{\rm th} = f_{\rm th} \tau^{\rm Ohm}_{\rm c}$, and
$\tau_{\rm cci}=f_{\rm cci} \tau^{\rm Ohm}_{\rm c}$, respectively.}
\begin{equation}
\tau^{\rm Ohm}_{\rm c} \equiv \frac{1}{\mu \omega_{\rm c}}.
\end{equation}
With respect to the current-carrier inertia terms, putting aside the detail of
the factor $f_{\rm cci}$, we have the characteristic
time scale 
\begin{equation}
\tau_{\rm cci} \equiv \frac{1}{\sqrt{\mu} \omega_{\rm c}}
= \sqrt{\mu} \tau^{\rm Ohm}_{\rm c}.
\end{equation}
It is noted that $\tau^{\rm Ohm}_{\rm c}$ gives a more severe condition
of the non-ideal MHD effects than $\tau_{\rm cci}$ because of
$\tau_{\rm cci} = \sqrt{\mu} \tau^{\rm Ohm}_{\rm c} < \tau^{\rm Ohm}_{\rm c}$.

The characteristic length scales of the non-ideal MHD effects are
\footnote{The more precise characteristic length scale of resistive, Hall, 
thermo-electromotive force, and current-carrier inertia effects are given by 
$L_{\rm r} = f_{\rm r} L^{\rm Ohm}_{\rm c}$,
$L_{\rm H} = f_{\rm H} L^{\rm Ohm}_{\rm c}$, 
$L_{\rm th} = f_{\rm th} L^{\rm Ohm}_{\rm c}$, 
and $L_{\rm cci}=f_{\rm cci} L^{\rm Ohm}_{\rm c}$, respectively.}
\begin{eqnarray}
L^{\rm Ohm}_{\rm c} &=& u_{\rm A} \tau^{\rm Ohm}_{\rm c} 
= \frac{u_{\rm A}}{\mu \omega_{\rm c}}
= \frac{c}{\sqrt{\mu} \omega_{\rm p}}
= \frac{1}{\sqrt{\mu}} l_{\rm s}, \\
L_{\rm cci} &=& u_{\rm A} \tau_{\rm cci} 
= \frac{u_{\rm A}}{\sqrt{\mu} \omega_{\rm c}} 
= \frac{c}{\omega_{\rm p}} 
=l_{\rm s},
\end{eqnarray}
where $\displaystyle l_{\rm s} = \frac{c}{\omega_{\rm p}}$ is the plasma skin depth.
Here, we also have 
$L_{\rm cci} = \sqrt{\mu} L^{\rm Ohm}_{\rm c} < L^{\rm Ohm}_{\rm c}$ and
$L^{\rm Ohm}_{\rm c}$ gives a more severe condition
of the non-ideal MHD effects than $L_{\rm cci}$.
Then, to confirm that non-ideal MHD terms are all negligible,
we just investigate whether both $\tau^{\rm Ohm}_{\rm c}$ and $L^{\rm Ohm}_{\rm c}$ 
are much smaller than
the time and length scales of the plasma phenomena $\tau$ and $L$ 
($\tau \gg \tau^{\rm Ohm}_{\rm c}$, $L \gg L^{\rm Ohm}_{\rm c}$)
because the factors ($f_{\rm r}, f_{\rm H}, f_{\rm th}, f_{\rm cci}$) are usually 
equal to or less than unity.
When $\tau^{\rm Ohm}_{\rm c} \gtrsim \tau$ or $L^{\rm Ohm}_{\rm c} \gtrsim L$,
it is possible that $\tau_{\rm cci} \ll \tau$ or $L_{\rm cci} \ll L$ for normal plasma
because $\sqrt{\mu} = 1/\sqrt{1836} = 1/42.8$ for normal plasma.
In such a case, we neglect the current-carrier inertia terms of the generalized Ohm law
and momentum equations, while the resistive term, Hall effect, and thermo-electromotive
force are significant.
Otherwise, that is, $\tau \lesssim \tau_{\rm cci}$ and $L \lesssim L_{\rm cci}$,
all of the non-ideal MHD effects are significant if the factors 
are nearly equal to unity ($f_{\rm r}, f_{\rm H}, f_{\rm th}, f_{\rm cci} \sim 1$).

It is noted that the finite resistivity may cause 
the magnetic reconnection and drastic phenomena even if the term is
much smaller than the term of $\VEC{U} \times \VEC{B}$.
Then, we have to take into account the electric resistive term even if $S_{\rm M}$ 
is much larger than unity. Furthermore, it is noted that current-carrier inertia also may
cause the magnetic reconnection \citep{hirota13,hirota15}.

For estimation of the variables of plasmas around these black holes, we have to give the
black hole mass $M_{\rm BH}$, plasma density $\rho$, and magnetic field $B$ of the plasma.
We had no direct observation that determines the variables around any black holes
before the observation of the EHT Collaboration \citep{eht19a,eht19b}.
Now, we employ the data set of $M_{\rm BH}$, accretion rate $\dot{M}$, 
$\rho$, temperature $T$, and $B$ 
in the accretion disks of M87* from the EHT Collaboration results.
Using the EHT observation data,
we check the characteristic scales of non-ideal MHD effects, 
$\tau_{\rm c}^{\rm Ohm} = 1/(\mu \omega_{\rm c})$ 
and $L_{\rm c}^{\rm Ohm} = l_{\rm s}/\sqrt{\mu}
= c/(\sqrt{\mu} \omega_{\rm p})$ on the plasma in M87*.
Here, we assume the plasma observed by EHT is normal plasma
($\Delta \mu =0$ and $\mu = 1/1836$).
\citet{eht19b} reported that the observation of M87* is explained by a simple,
spherical, one-zone model for the source 
as
\begin{eqnarray}
n_{\rm e} &=& 2.9 \times 10^4 \left ( \frac{r}{5 r_{\rm g}} \right )^{-1.3}
\beta_{\rm p}^{0.62} \left ( \frac{T_{\rm i}}{3 T_{\rm e}} \right )^{-0.47} \, [{\rm cm^{-3}}], \label{n2simplemodel2eht} \\
B &=& 4.9 \left ( \frac{r}{5 r_{\rm g}} \right )^{-0.63}
\beta_{\rm p}^{-0.19} \left ( \frac{T_{\rm i}}{3 T_{\rm e}} \right )^{-0.14} \, [{\rm G}],\label{b2simplemodel2eht}\\
T_{\rm i} &=& 0.202 \times 10^{12} 
\left ( \frac{r}{5 r_{\rm g}} \right ) \, [{\rm K}].\label{t2simplemodel2eht}
\end{eqnarray}

First, we evaluate the plasma parameters at $r=5r_{\rm g}$ in the model, for example.
At $r=5 r_{\rm g}$, we have $f=\rho/h \approx 1$, $\omega_{\rm c} \approx
\omega_{\rm c}^{\rm prm} = 8.6 \times 10^7 {\rm s^{-1}}$, 
$u_{\rm A} \approx v_{\rm A}^{\rm prm} = 2.8 \times 10^9 {\rm cm \, s^{-1}}$, 
$\omega_{\rm p} \approx \omega_{\rm p}^{\rm prm} = 9.6 \times 10^6 {\rm s^{-1}}$, 
$c_{\rm s} \approx c_{\rm c}^{\rm prm} = 4.9 \times 10^9 {\rm cm \, s^{-1}}$, 
and $\nu_{\rm ei} = 1.2 \times 10^{-12} {\rm s^{-1}}$.
Then, we found characteristic scales of the non-ideal MHD phenomena as
$\tau^{\rm Ohm}_{\rm c} = 1/ (\mu \omega_{\rm c}) = 5.0 \times 10^{-5}$ s and
$L^{\rm Ohm}_{\rm c}=1.3 \times 10^3$m.
Considering the spatial dependence on 
characteristic scales:
$\tau^{\rm Ohm}_{\rm c} = 1/(\mu \omega_{\rm c}) \propto 
B^{-1} f_{\rm rel}^{-1} \propto r^{6.3} \beta_{\rm p}^{0.11} 
( T_{\rm i}/T_{\rm e} )^{0.14} f_{\rm rel}^{-1} $, 
$L^{\rm Ohm}_{\rm c} = c/(\sqrt{\mu} \omega_{\rm p})  \propto 
n^{-1/2} f_{\rm rel}^{-1/2} \propto r^{0.65} \beta_{\rm p}^{-0.31} 
( T_{\rm i}/T_{\rm e} )^{0.24} f_{\rm rel}^{-1/2}$, we have
\begin{eqnarray}
\tau^{\rm Ohm}_{\rm c} & = & \frac{1}{\mu \omega_{\rm c}} = 5.0 \times 10^{-5}
\left ( \frac{r}{5 r_{\rm g}} \right )^{063} \beta_{\rm p}^{0.11} 
\left ( \frac{T_{\rm i}}{T_{\rm e}} \right )^{0.14} f_{\rm rel}^{-1} \, [{\rm s}], \\ 
L^{\rm Ohm}_{\rm c} & = & \frac{c}{\sqrt{\mu} \omega_{\rm p}} =  1.3 \times 10^3
\left ( \frac{r}{5 r_{\rm g}} \right )^{0.65}  
\left ( \frac{T_{\rm i}}{T_{\rm e}} \right )^{0.24} f_{\rm rel}^{-1/2}  \, [{\rm m}].
\end{eqnarray}
These scales become larger as the outer region becomes farther from the black hole.
%

Incidentally, we note that the extended Debye length of the plasma is 
$\displaystyle \lambda_{\rm D} = \frac{1}{\sqrt{\mu}}
\frac{c_{\rm s}^0}{\omega_{\rm p}} 
= \sqrt{(1 - 2 \mu) \beta_{\rm p}} \frac{u_{\rm A}}{\sqrt{\mu} \omega_{\rm p}}
\sim 8.4$ cm where we calculate $\beta_{\rm p} \sim 0.84$ at $r = 5 r_{\rm g}$.
The particle number in the Debye sphere of a charged particle,
$\displaystyle N_{\rm D} = \frac{4 \pi}{3} \lambda_{\rm D}^3 n
= 7 \times 10^7$, is much greater than unity, and the plasma has collective 
property as a plasma.

As an example of the minimum of the characteristic scales of phenomena of plasmas
($L$ and $\tau$) around the black holes, we consider the current sheet
that causes the magnetic reconnection in the accretion disk around the black hole.
The minimum scales of the magnetic reconnection are roughly estimated by the thickness
of the current sheet $L_{\rm CS}$, which is calculated by the minimum scale
of the magnetorotational instability (MRI),
$\displaystyle L_{\rm CS} \sim \lambda_{\rm MRI} = 
4 \sqrt{\frac{2}{3}} \frac{v_{\rm A}}{\Omega}$
(see Chapter 8 in \citet{shibata99} or Chapter 4 in \citet{tajima02}).
Here, $\Omega=\sqrt{GM_{\rm BH}/r^3}=c/\sqrt{2} r_{\rm S} (r/r_{\rm S})^{-3/2}$ 
is the angular velocity of the disk and $v_{\rm A} = \sqrt{B^2/ \mu_0 \rho}$ 
is the Alfven velocity; $\Omega \approx \sqrt{GM_{\rm BH}/r^3} > c/r_{\rm S}$.
The Alfven transit time of the current sheet is given by $\tau_{\rm A}=L_{\rm CS}/v_{\rm A}
=8/\sqrt{3} \tau_{\rm S} (r/r_{\rm S})^{3/2}$, where
$\tau_{\rm S}=r_{\rm S}/c$ is the Schwarzschild transit time.
From the M87* observation by EHT, we have $M_{\rm BH} = 6.5 \times 10^9 M_\odot$,
$r_{\rm S} = 1.9 \times 10^{15}$ cm, and
$\tau_{\rm S} = 6.3 \times 10^4$ s.
The thickness of the current sheet is calculated as 
\begin{equation}
L_{\rm CS} 
= \frac{8}{\sqrt{3}} \left (\frac{r}{r_{\rm S}} \right )^{3/2} \frac{v_{\rm A}}{c} r_{\rm S}
= 7.5 \times 10^{15} \left (\frac{r}{5 r_{\rm g}} \right )^{1.52} \rm [cm].
\end{equation}
The values of the spatial and temporal scales, $L_{\rm CS}$ and $\tau_{\rm A}$ 
are much larger than the critical variables, $\tau^{\rm Ohm}_{\rm c}
= \tau^{\rm Ohm}_{\rm cci}/\sqrt{\mu}$ and 
$L^{\rm Ohm}_{\rm c} = L^{\rm Ohm}_{\rm cci}/\sqrt{\mu}$,
respectively. This suggests the validity of the 
resistive GRMHD equations in the phenomena in 
the reconnection regions around the black holes.


\section{Summary \label{secsd}}

In this paper, we determined the $\Theta$ term of the generalized relativistic
Ohm law, which had not been determined in our previous works
\citep{koide08,koide09,koide10}.
We have now obtained an explicitly closed \textcolor{black}{system} of generalized GRMHD equations and
evaluated the terms of the non-ideal MHD effects in these 
equations (Equations (\ref{mdfgrmmom}) and (\ref{mdfgrmohm})). 
There are two main characteristic scales of the non-ideal MHD 
effects with respect to time and
length. These scales come from 
the generalized Ohm law (\ref{mdfgrmohm})
and are given by $\tau^{\rm Ohm}_{\rm c} = 1/(\mu \omega_{\rm c})$
and $L^{\rm Ohm}_{\rm c} = c/(\sqrt{\mu} \omega_{\rm p})$.
In more detail, the additional characteristic time and length
scales come from current-carrier inertia both in the generalized Ohm law and in the momentum
equation and are given by 
$\tau_{\rm cci} = \sqrt{\mu} \tau^{\rm Ohm}_{\rm c}$ and 
$L_{\rm cci} = \sqrt{\mu} L^{\rm Ohm}_{\rm c}$, which are smaller
by a factor $\sqrt{\mu} \le 1/2$ than $\tau^{\rm Ohm}_{\rm c}$
and $L^{\rm Ohm}_{\rm c}$, respectively.
We evaluated the additional terms of the generalized relativistic Ohm law
with the plasma parameters ($T_{\rm i}$, $n_{\rm e}$, and $B$) 
obtained by EHT observations of M87* and
found that the additional terms of resistive electromotive force, the Hall effect,
thermo-electromotive force, and the current-carrier (electron) inertia effect
are negligible compared to the $\VEC{U} \times \VEC{B}$ term
of the generalized relativistic Ohm law 
for global-scale phenomena around the black hole, whose characteristic
length scale is given by $L_{\rm CS} \sim r_{\rm S}$.

While the resistive term is negligible in the global phenomena around the black hole,
the magnetic reconnection has been suggested to occur frequently
in black-hole magnetospheres by a number of
ideal GRMHD simulations \citep[e.g.,][]{koide00,koide06,mckinney06}. 
However, it should be emphasized that
magnetic reconnection in the ideal GRMHD simulations is caused by numerical
resistivity; this resistivity often results in a fatal error for the numerical results
and should be avoided. 
To perform GRMHD simulations of magnetic reconnection around a black hole
without numerical resistivity,
we must develop a highly accurate resistive GRMHD code.
Recently, \citet{indako19} performed resistive GRMHD simulations 
of the magnetic reconnection
around a black hole with a simple magnetic configuration and relativity
small Reynolds number ($S_{\rm M} \sim 10^4$).
This work was the first resistive GRMHD simulation of magnetic reconnection around
a black hole; however, the Reynolds number was not sufficiently large 
and the magnetic configuration
was too simple to apply the results to astrophysical objects.
To perform simulations with a suitably high magnetic Reynolds number and
a magnetic configuration appropriate for the astrophysical situation,
we require the advanced numerical technique of an implicit method with high
accuracy \citep{bucciantini13,tomei20}.

On the other hand, explosive magnetic reconnection in a collisionless
plasma has been proposed actually using the generalized Ohm law
\citep{hirota13,hirota15}.
The current-carrier inertia (electron inertia) term plays an important role
in the explosive magnetic reconnection model. 
\citet{comisso14} found similar magnetic reconnection
with the generalized relativistic MHD equations given by \citet{koide09}.
The normalized reconnection rate was given by
\begin{equation}
\frac{v_{\rm in}}{c} \sim \sqrt{\frac{1}{S} + \frac{1}{4f} \frac{l_{\rm s}}{L}},
\end{equation}
where $v_{\rm in}$ is the inflow velocity of plasma toward the reconnection region.
The first term in the square root on the right-hand side represents the 
magnetic-reconnection rate of the Sweet-Parker reconnection model, in which
electric resistivity causes the reconnection.
The second term in the square root represents the rate of the reconnection
due to current-carrier (electron) inertia.
The ratio of the first and second terms for the relativistic
magnetic reconnection ($u_{\rm A} \sim c$) is calculated as
\begin{equation}
\displaystyle \frac{\frac{1}{4f} \frac{l_{\rm s}}{L}}{\frac{1}{S}}
= \frac{\mu_0 u_{\rm A} l_{\rm s}}{4 f \eta}
= \frac{\mu_0 u_{\rm A} c}{4 f \omega_{\rm c}} 
\frac{2 \omega_{\rm p}}{\mu_0 c^2 \nu_{+-}}
\frac{\mu m}{m_{\rm e}}
= \frac{u_{\rm A}}{2 f c} \frac{\mu m}{m_{\rm e}} \frac{\omega_{\rm p}}{\nu_{+-}}
\sim \frac{u_{\rm A}}{c} \frac{\omega_{\rm p}}{\nu_{+-}} \gg 1.
\end{equation}
The magnetic reconnection due to the effect of current-carrier inertia
would be significant compared to resistive magnetic reconnection.
To confirm this explosive reconnection, a numerical simulation
of the generalized GRMHD is required.

In this paper, we used a simple 1-D model 
(Equations (\ref{n2simplemodel2eht})--(\ref{t2simplemodel2eht})) 
based on the EHT observation
of M87* \citep{eht19b} to evaluate the significance of the non-ideal MHD effect in
the surrounding plasma. Using this simple model, we concluded that
the ideal GRMHD approximation works well for the global phenomena of
plasma \textcolor{black}{around} M87*.
However, the plasmas around M87* are actually complex,
since the region is composed of the torus, the accretion disk, the corona, 
the outflow (wind), and the jet.
The 1-D model is too simplistic to grasp the detailed plasma behavior
around the black hole. We must improve the significance evaluation of the non-ideal MHD
effect based on a forthcoming, more actual model taken from EHT observations of
not only M87*, but also of Sgr A*. It is worth continuing to check the significance
of the non-ideal MHD effect, because it would change the plasma dynamics
drastically from the results of ideal GRMHD simulations.

Numerical simulation of the generalized GRMHD is necessary to confirm and reveal
the specific phenomena caused by non-ideal MHD effects.
The EoSs (74) and (75) with respect to $h$ and $\Delta h$ 
in \citet{koide09} provide closure to 
the generalized GRMHD equations (\ref{mdfgrmnum})-(\ref{etarhoetheta2}).
Such \textcolor{black}{numerical} calculation is possible in principle,
although it becomes drastically difficult compared to the ideal
GRMHD simulations. This is because we have to treat the displacement
current $\partial \VEC{E}/\partial t$ in Ampere's law and the inertia of
the current density $[\mu m/(ne^2)] \partial [\{h^\ddagger/(nm)\} \VEC{J}]/\partial t$ 
in the generalized Ohm law \textcolor{black}{explicitly}.
In the ideal GRMHD calculations, the former is implicitly accounted for
and the latter can be  neglected entirely.
Furthermore, we have to consider the zeroth component of Ohm law
to calculate the enthalpy-density difference, $\Delta h$, of relativistically hot plasmas 
around the black hole.
Thus, appropriate simplifications of the generalized GRMHD equations, 
especially of Ohm law, are required for adequate 
numerical study.
\textcolor{black}{For this simplification, the characteristic scales of the non-ideal MHD
phenomena will provide a basic guide. The adjusted closed system of the generalized
GRMHD equations will play a significant role in forthcoming numerical
simulations of magnetized plasmas around the black hole in the new era with
the EHT observations.}

\acknowledgments

I am grateful to Mika Koide and Shohei Sakai for their helpful comments on this paper.



\appendix

\section{Derivation of a 3+1 formalism of divergence of symmetric tensor}

We derive a 3 + 1 formalism of divergence of symmetric tensor, 
$\nabla_\nu T^{\mu\nu} = F^{\mu}$.

The 4-acceleration of the normal frame is given by $a_\mu = N_{\mu ; \nu} N^\nu$
and we have
\begin{equation}
a_\mu = ( - \alpha N^k (\ln \alpha )_{,k}, (\ln \alpha)_{,i}), \verb!   !
a^\mu = ( 0, \gamma^{ik} ( \ln \alpha)_{,k}), 
\end{equation}
where the subscript ``," denotes the partial derivative $\partial/\partial x^i$.
Using the projection tensor to
the time constant hypersurface $\mathcal{P}^{\alpha \beta} = g^{\alpha \beta}
+N^\alpha N^\beta$, we define the extrinsic curvature tensor by
$K_{ij} \equiv - \mathcal{P}^\mu_i \mathcal{P}^\nu_j N_{\mu;\nu}= - N_{i;j}$,
where the subscript ``;" denotes the covariant derivative $\nabla_i$.
In the stationary spacetime ($\gamma_{ij,t}=0$), we have 
\begin{equation}
K_{ij} = - \frac{1}{2 \alpha} [\gamma_{ij,t} + \nabla_i (\alpha N_j) + 
\nabla_j (\alpha N_i)]
= - \frac{1}{2 \alpha} [\gamma_{kj} (\alpha N^k)_{,i} + \gamma_{ik} (\alpha N^k)_{,j}
+ \gamma_{ij,k} \alpha N^k] .
\end{equation}

Using the normal vector of the hypersurface of time constant $N^\mu$ and 
projection tensor toward the hypersurface $\mathcal{P}^{\mu\nu}$, we separate 
the 4-vector $F^\mu$ into temporal and spatial components:
\begin{eqnarray}
& \tilde{F}^{\dagger} \equiv - F^\mu N_\mu
= - T^{\mu \nu}_{; \nu} N_\mu, 
\label{conlawenem}\\
& \tilde{F}_i \equiv  F^\mu \mathcal{P}_{i\mu}
= {T^{\mu \nu}}_{; \nu} \mathcal{P}_{i \mu} .
\label{conlawmoem}
\end{eqnarray}
Note that $\tilde{F}^\dagger$ and $\tilde{F}^\mu$ reproduce $F^\mu$ as
\begin{equation}
F^\mu = \tilde{F}^\dagger N^\mu + \tilde{F}^\mu.
\end{equation}
Equation (\ref{conlawenem}) yields a scalar-like equation 
such as the energy conservation law,
and Equation (\ref{conlawmoem}) yields a 3-vector conservation equation 
such as the momentum conservation
law.
Similarly, when we separate $T^{\mu\nu}$ into
\begin{eqnarray}
\tilde{u} & = T^{\rho \sigma} N_\rho N_\sigma , \\
\tilde{S}_\mu & = - T^{\rho \sigma} \mathcal{P}_{\mu\rho} N_\sigma , \\
\tilde{T}_{\mu \nu} & = T^{\rho \sigma} \mathcal{P}_{\mu\rho}
\mathcal{P}_{\nu\sigma} .
\end{eqnarray}
Here, we found 
\begin{equation}
T^{\mu\nu} = \tilde{u} N^\mu N^\nu +
\tilde{S}^{\mu} N^\nu + N^\mu \tilde{S}^{\nu}+ \tilde{T}^{\mu\nu},
\label{bunkai}
\end{equation}
and $\tilde{u}=T^{\tilde{0}\tilde{0}}$,
$\tilde{S}_i = T^{\tilde{0}}_{\tilde{i}}$,
$\tilde{T}_{ij} = T_{\tilde{i}\tilde{j}}$,
$\tilde{F}^{\dagger} = F^{\tilde{0}}$,
$\tilde{F}_{i} = F_{\tilde{i}}$.
Equation (\ref{conlawenem}) is written by
\begin{equation}
\tilde{F}^\dagger = - (T^{\mu\nu} N_\mu)_{;\nu} + T^{\mu\nu} N_{\mu;\nu}
= - \frac{1}{\sqrt{-g}} (\sqrt{-g} T^{\mu\nu} N_\nu)_{,\mu} + T^{\mu\nu} N_{\mu;\nu}.
\end{equation}
Using $a_\mu = N_{\mu;\nu} N^\nu$, $K_{ij}=-N_{i;j}$, $\sqrt{-g}=\alpha \sqrt{\gamma}$,
and Equation (\ref{bunkai}), we have
\begin{equation}
\tilde{F}^\dagger = \frac{1}{\alpha \sqrt{\gamma}} (\sqrt{\gamma} \tilde{u})_{,t}
+  \frac{1}{\alpha \sqrt{\gamma}} [\alpha \sqrt{\gamma} (\tilde{S}^k + N^k \tilde{u})]_{,k}
+ (\ln \alpha)_{,i} \tilde{S}^i - K_{ij} \tilde{T}^{ij}.
\end{equation}
Multiplying by $\alpha$, we obtain
\begin{equation}
\alpha \tilde{F}^{\dagger} = \frac{1}{\sqrt{\gamma}} \frac{\partial}{\partial t} 
(\sqrt{\gamma} \tilde{u})
+ \frac{1}{\sqrt{\gamma}} \frac{\partial}{\partial x^k}
[\alpha \sqrt{\gamma} (\tilde{S}^k + \tilde{u} {N}^k)]
+ \frac{\partial \alpha}{\partial x^i} \tilde{S}^i
- K_{ij} \tilde{T}^{ij} .
\label{3+1enem6ch3}
\end{equation}
When we use $\displaystyle \alpha K_{ij} \tilde{T}^{ij}= \left [\gamma_{jk} \frac{\partial }{\partial x^i} (\alpha N^k)                            
+  \frac{1}{2} \alpha N^k  \frac{\partial \gamma_{ij}}{\partial x^k}                        
\right ] \tilde{T}^{ij}$, we get
\begin{eqnarray}
& \alpha \tilde{F}^{\dagger} = \displaystyle \frac{\partial}{\partial t} \tilde{u}
+ \frac{1}{\sqrt{\gamma}} \frac{\partial}{\partial x^i}
[\alpha \sqrt{\gamma} (\tilde{S}^i + {N}^i \tilde{u})]
+ \frac{\partial \alpha}{\partial x^i} \tilde{S}^i
\nonumber \\
& \displaystyle + \left [ \gamma_{jk} \frac{\partial}{\partial x^i} (\alpha N^k)
+ \frac{1}{2} \alpha N^k \frac{\partial  \gamma_{ij}}{\partial x^k}
\right ] \tilde{T}^{ij}.
\label{3+1enem}
\end{eqnarray}
With respect to Equation (\ref{conlawmoem}), we have
\begin{equation}
\tilde{F}_i = T^{\nu}_{i;\nu} = \frac{1}{\sqrt{-g}} (\sqrt{-g} T^{\nu}_i)_{,\nu}
- \frac{1}{2} g_{\alpha\beta , i} T^{\alpha \beta}.
\end{equation}
Using $a_\mu = N_{\mu;\nu} N^\nu$, $\sqrt{-g}=\alpha \sqrt{\gamma}$,
and Equation (\ref{bunkai}), we get
\begin{equation}
\tilde{F}_i = \frac{1}{\alpha \sqrt{\gamma}} ( \sqrt{\gamma} \tilde{S}_i)_{,t}
+ \frac{1}{\alpha \sqrt{\gamma}} [\alpha \sqrt{\gamma} 
(\tilde{T}^k_i + N^k S_i)]_{,k} + \tilde{u} (\ln \alpha)_{,i}
+ \frac{1}{\alpha} (\alpha N^k)_{,i} \tilde{S}_k - \frac{1}{2} \gamma_{kl,i} \tilde{T}^{kl}.
\end{equation}
Multiplying by $\alpha$, we obtain
\begin{equation}
 \alpha \tilde{F}_i =
\frac{\partial}{\partial t} \tilde{S}_i
+ \frac{1}{\sqrt{\gamma}} \frac{\partial}{\partial x^k}
[ \alpha \sqrt{\gamma} (\tilde{{T}^k}_i + {N}^k \tilde{S}_i) ]
+ \frac{\partial \alpha}{\partial x^i} \tilde{u}
 + \frac{\partial}{\partial x^i} (\alpha N^k) \tilde{S}_k
- \frac{\alpha}{2} \frac{\partial \gamma_{jk}}{\partial x^i} \tilde{T}^{jk}
.
\label{3+1moem}
\end{equation}
Furthermore, using a 3-covariant derivative, which is given by
$\displaystyle ^{(3)}\nabla_k A^k_i = \frac{1}{\sqrt{\gamma}}              
\partial_k (\sqrt{\gamma} A^k_i) - \, ^{(3)}\Gamma^k_{ij} A^j_k$
for a 3-tensor $A^k_i$, we have 
\begin{equation}
\alpha \tilde{F}_i =
\frac{\partial}{\partial t} \tilde{S}_i
+ \, ^{(3)} \nabla_k [ \alpha (\tilde{{T}^k}_i +  \tilde{S}_i {N}^k) ]
+ \tilde{u} \frac{\partial \alpha}{\partial x^i}
- \tilde{S}_k \, ^{(3)}\nabla_i (\alpha N^k) .
\label{3+1moemch3}
\end{equation}

\end{document}